\documentclass[sigconf,screen]{acmart}
\usepackage{array}
\usepackage{tikz}
\usepackage{amsmath}
\usepackage{amsfonts}
\usepackage{textcomp}
\usepackage{xcolor}
\usepackage{url}

\usepackage{afterpage}
\usepackage{tabularx}
\usepackage{algorithm}
\usepackage{algorithmic}
\usepackage{graphicx}
\usepackage{subfigure} 
\usepackage{fancybox}
\usepackage{xspace}
\usepackage{threeparttable}
\usepackage{multirow}
\usepackage{enumitem}
\usepackage{seqsplit}
\usepackage{appendix}
\usepackage{booktabs}
\usepackage{varwidth}
\usepackage{tcolorbox}
\usepackage{shadowtext}
\usepackage{soul} 
\usepackage{makecell}
\usepackage{cellspace}
\definecolor{green2}{RGB}{0,125,0}

\sethlcolor{lightgray!30}

\newcommand{\one}{({\em i}\/) }
\newcommand{\two}{({\em ii}\/) }
\newcommand{\three}{({\em iii}\/) }
\newcommand{\four}{({\em iv}\/)\xspace}

\setcopyright{none}
\renewcommand\footnotetextcopyrightpermission[1]{}

\date{}

\begin{document}
\title{Understanding User Privacy Perceptions of GenAI Smartphones}


\author{Ran Jin}
\orcid{0009-0002-5052-9458}
\authornotemark[1]
\affiliation{
  \institution{Huazhong University of Science and Technology}
  \city{Wuhan}
  \country{China}
}
\email{jran@hust.edu.cn}

\author{Liu Wang}
\orcid{0000-0002-3982-4993}
\authornote{Ran Jin and Liu Wang are the co-first authors.}
\affiliation{%
  \institution{Huazhong University of Science and Technology}
  \city{Wuhan}
  \country{China}
}
\email{liuwang@hust.edu.cn}

\author{Shidong Pan}
\orcid{0000-0002-2162-0407}
\authornotemark[2]
\affiliation{%
  \institution{New York University}
  \city{New York}
  \country{USA}
}
\email{Shidong.Pan@nyu.edu}

\author{Luona Xu}
\orcid{0009-0009-3219-1108}
\affiliation{%
  \institution{Huazhong University of Science and Technology}
  \city{Wuhan}
  \country{China}
}
\email{u202212106@hust.edu.cn}

\author{Tianming Liu}
\orcid{0000-0002-5216-933X}
\authornote{Shidong Pan and Tianming Liu are the corresponding authors.}
\affiliation{%
  \institution{Huazhong University of Science and Technology}
  \city{Wuhan}
  \country{China}
}
\email{tmliu@hust.edu.cn}

\author{Haoyu Wang}
\orcid{0000-0003-1100-8633}
\affiliation{%
  \institution{Huazhong University of Science and Technology}
  \city{Wuhan}
  \country{China}
}
\email{haoyuwang@hust.edu.cn}

\begin{abstract}

GenAI smartphones, which natively embed generative AI at the system level, are transforming mobile interactions by automating a wide range of tasks and executing UI actions on behalf of users.
Their superior capabilities rely on continuous access to sensitive and context-rich data, raising privacy concerns that surpass those of traditional mobile devices. 
Yet, little is known about how users perceive the privacy implications of such devices or what safeguards they expect, which is especially critical at this early stage of GenAI smartphone adoption.
To address this gap, we conduct 22 semi-structured interviews with everyday mobile users to explore their usage of GenAI smartphones, privacy concerns, and privacy design expectations. 
Our findings show that users engage with GenAI smartphones with limited understanding of how these systems operate to deliver functions, but show heightened privacy concerns once exposed to the technical details. 
Participants' concerns span the entire data lifecycle, including nontransparent collection, insecure storage, and weak data control.
In a follow-up focus group, participants discuss a range of privacy-enhancing suggestions that call for coordinated changes across system-level controls, data management practices, and user-facing transparency.
Their concerns and suggestions offer user-centered guidances for designing GenAI smartphones that balance functionality with privacy protection, offering valuable takeaways for system designers and regulators.
\end{abstract}



\maketitle

\section{Introduction}
\begin{figure}[t]
    \centering
    \includegraphics[width=\linewidth]{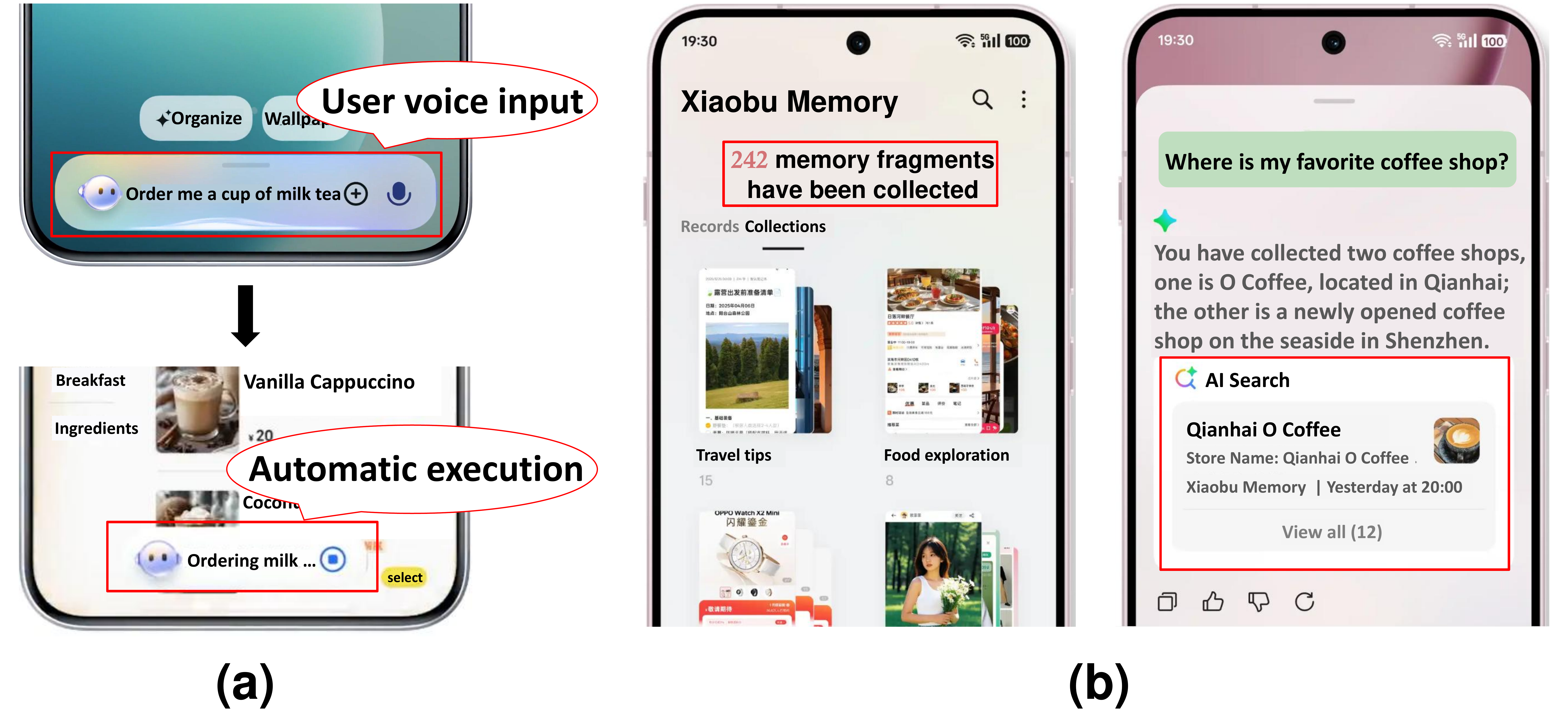}
    \caption{GenAI smartphone usage scenarios. (a) Autonomous task completion: GenAI autonomously plans and executes multi-step actions across apps; (b) Contextual memory: GenAI persistently records and structures contextual information for future querying.}
    \label{fig:usage scenarios}
    \vspace{-0.15in}
\end{figure}

The rapid advancement of Generative Artificial Intelligence (GenAI) is fundamentally reshaping how people interact with digital technologies~\cite{Definition_of_GenAI, kalota2024aprimer, mckinsey2023generative_ai}.
GenAI encompasses machine learning models capable of generating novel, contextually relevant text, images, audio, and other media by learning complex patterns from massive, diverse datasets.
As GenAI technologies continue to evolve, they are increasingly being embedded into ubiquitous devices–most notably, the smartphones.
According to International Data Corporation (IDC)~\cite{GenAI_phones}, 
\textbf{GenAI smartphones} are defined as devices featuring a system-on-a-chip (SoC) capable of running on-device Generative (GenAI) models more quickly and efficiently leveraging a neural processing unit.
These devices either host generative models locally or connect to cloud-based services, enabling system-level integration of GenAI. Powered by these models, the devices provide users with advanced capabilities, such as autonomous task completion and contextual memory (as shown in \autoref{fig:usage scenarios}). 
Additionally, these GenAI smartphones can function as proactive digital assistants~\cite{chaudhary2024ai}, learning from individual users’ habits and preferences over time to anticipate needs and make everyday activities faster, smoother, and more intuitive~\cite{ apple_intelligence, motorola}.
Major smartphone manufacturers such as HONOR~\cite{HONOR}, Apple~\cite{Apple}, OPPO~\cite{OPPO}, and Xiaomi~\cite{Xiaomi}, have begun integrating GenAI capabilities into their latest devices, signaling a significant shift toward AI-native mobile experiences. 
IDC forecasts that between 2024 and 2028, the GenAI smartphone market will maintain a compound annual growth rate of 78.4\%, with shipments expected to reach 912 million units by 2028~\cite{GenAI_phones}. 

However, the deep and pervasive integration of GenAI into mobile devices introduces significant privacy challenges. 
To support context-aware reasoning and automated task execution, GenAI smartphones often require continuous access to highly sensitive and context-rich information, such as on-screen content, voice interactions, personal documents, and sensor readings~\cite{GenAI_phones, wu2024first}. 
These capabilities fundamentally expand the scope of data access on mobile devices.
At the system level, 
GenAI smartphones often operate across apps by relying on elevated system permissions (e.g., accessibility services), which may enable unintended permission escalation and weaken existing user consent boundaries~\cite{pan2025lookprivacyrisksandroid}.
At the model level, on-device GenAI models may be vulnerable to compromise, modification, or misuse, potentially exposing sensitive data processed locally~\cite{golda2024privacy, das2025security, yao2024survey}. 
In addition, GenAI smartphones may be vulnerable to errors or attacks due to intrinsic model vulnerabilities or imperfect system integration~\cite{liu2024generative, feretzakis2024privacy, xu2021privacy}, further compounding these privacy risks. 
At the same time, the powerful functionality and convenience offered by GenAI smartphones may encourage users to tolerate or overlook such risks in practice, intensifying the longstanding tension between perceived utility and privacy protection~\cite{Tran2025UnderstandingPN}. In this sense, GenAI smartphones may exacerbate the \textit{privacy paradox}~\cite{norberg2007privacy, acquisti2015privacy}, whereby users express privacy concerns yet continue to engage with data-intensive technologies in exchange for convenience and enhanced functionality.

While prior research has examined privacy concerns in mobile platforms~\cite{degirmenci2020mobile, saeed2024usable, maseeh2023exploring}, large language models (LLMs)~\cite{he2025emerged, ma2025privacy}, and emerging mobile GUI agents~\cite{zhang2024privacy, shao2024privacylens}, dedicated studies focusing on GenAI smartphones remain scarce due to their recent emergence as system-level AI-integrated devices.
Their distinctive capabilities (such as cross-app autonomy, continuous contextual awareness, and hybrid on-device and cloud processing) fundamentally reshape how personal data is accessed and used, introducing new privacy dynamics that are not yet well understood. Particularly, there is limited empirical understanding of how end users perceive privacy in the context of GenAI smartphone usage and what design directions they consider appropriate for protecting it. 
As GenAI smartphones are still at an early stage of adoption, it is especially important to understand users’ privacy concerns, expectations, and needs before design patterns and system behaviors become entrenched. Without such understanding, efforts to design privacy-conscious GenAI smartphones risk misaligning with user values, thereby undermining trust and long-term adoption.
 
To address this gap, this study aims to develop an in-depth, user-centered understanding of privacy perceptions surrounding GenAI smartphones. Specifically, it seeks to explore how individuals experience and evaluate these devices in everyday use, how privacy concerns arise in relation to their technical and functional characteristics, and how users envision privacy-preserving designs that maintain the benefits of GenAI functionality. To achieve these aims, the study is guided by the following research questions:

\textbf{RQ1}: What are users’ perceptions, usage patterns, and privacy-convenience 
trade-offs when interacting with GenAI smartphones?

\textbf{RQ2}: How do users’ privacy concerns manifest with respect to the underlying technical implementation and functional design features of GenAI smartphones?

\textbf{RQ3}: What strategies and design recommendations do users propose to mitigate privacy risks while preserving the benefits of GenAI smartphones?

To address these questions, we conducted semi-structured, one-to-one interviews with everyday mobile users (n = 22). 
These participants possessed certain technological proficiency and familiarity to meaningfully engage with GenAI smartphone features, thereby ensuring informed reflections on both usage experiences and privacy concerns.
Our study focused on smartphones with system-level GenAI integration, as this category represents the most advanced and privacy-sensitive form of GenAI deployment on mobile devices. By examining users’ experiences with such systems, we can capture privacy perceptions and expectations that may not surface in studies of standalone AI apps or traditional smartphones.
In summary, we report the following key insights:
\begin{itemize}[leftmargin=*]
    \item Users currently engage with GenAI smartphones infrequently and possess only a limited understanding of their capabilities. However, after gaining deeper insight into the technical details of their operations, many participants expressed heightened privacy concerns. 
    While participants differ in the degree to which they prioritize convenience or privacy, they widely acknowledged privacy as a critical issue.
    \item Participants’ privacy concerns with GenAI smartphones span the entire data lifecycle, encompassing opaque data collection, insecure sharing and storage, and limited mechanisms for exercising data rights.
    They also expressed concern over unpredictable AI inferences and model misoperations that may lead to unintended disclosures.
    Beyond these data-centric risks, participants highlighted systemic challenges such as insufficient transparency and weak regulatory enforcement, as well as technical vulnerabilities arising from permission abuse and adversarial attacks.
    \item Participants’ user-informed suggestions call for coordinated changes across \one system-level control mechanisms (e.g., constraining autonomous agent actions and rethinking permission frameworks), \two data lifecycle management (e.g., centralized data control and secure data processing), and \three user-facing transparency (e.g., making data access and agent behavior visible during interaction). Technically experienced participants broadly agreed with the overall direction, while noting that system-layer proposals remain difficult to operationalize under current mobile architectures, highlighting the need for stronger OS-level support.
\end{itemize}

\section{Background}
\label{sec:background}

Commercial GenAI smartphones have begun to enter the consumer market in recent years.
One notable example is ByteDance's \textit{Doubao Phone}~\cite{Doubao_phone}, which features deep, system-level integration with the Doubao AI assistant to support agentic task execution and enhanced productivity. Similarly, major manufacturers such as Apple (with Apple Intelligence on iPhone 16 series)~\cite{Apple}, Samsung (with Galaxy AI on Galaxy S25 series)~\cite{Samsung}, and Xiaomi (with its AI assistant Xiao Ai)~\cite{Xiaomi} have all launched GenAI-powered flagship devices. 
These developments signal a broader industry shift toward AI-native smartphone experiences, in which generative models are embedded as core components of the mobile operating system.

\subsection{GenAI Smartphones}
\label{sec:GenAI smartphone}
GenAI refers to large machine learning models that are trained from large-scale datasets to generate contextually relevant content across modalities such as text, images, audio, and video~\cite{Definition_of_GenAI, mckinsey2023generative_ai, kalota2024aprimer, GozaloBrizuela2023ASO, Sengar2024GenerativeAI}. 
In recent years, GenAI has been increasingly integrated into mobile devices, giving rise to the concept of the \textit{GenAI smartphones}~\cite{GenAI_phones, yankodesign_genai_phones}. 
These devices embed GenAI models, either on-device or accessed via the cloud, directly into their operating systems.
This \textit{system-level integration} allows GenAI to leverage system-wide capabilities such as accessing screen content, system state, and coordinating actions across apps. As a result, GenAI smartphones can exhibit more autonomous and adaptive behaviors than app-bounded AI approaches such as mobile GUI agents.
From a hardware perspective, GenAI smartphones are typically equipped with system-on-chip (SoC) processors that incorporate dedicated neural processing units (NPUs),
enabling efficient execution of on-device GenAI models. 
At the software level, however, GenAI smartphones largely inherit the existing mobile permission framework, such as the Android permission model, which was originally designed for app-centric, user-initiated interactions. Under this framework, permissions are typically granted to individual apps to access specific resources (e.g., camera, microphone, location), with limited support for long-running, cross-app, or autonomous behaviors.
To enable agent-like capabilities under this legacy model, GenAI systems often rely on elevated system permissions, most notably \textit{accessibility services}, which were originally introduced to assist users with disabilities by allowing authorized services to observe screen content, detect UI events, and perform actions such as clicks and text input. In GenAI smartphones, these permissions are increasingly repurposed to support system-wide perception and automated interaction.

We can broadly categorize the functional scenarios of GenAI smartphones into two types. The first type consists of \textbf{direct data processing tasks}, such as image enhancement or text generation, which are typically completed in a single step and do not require sustained interaction with other system components.
The second type involves \textbf{more complex, agentic tasks} that require the integration of richer contextual information or simulation of user behavior through autonomous control of apps. For instance, in the task \textit{"Order me a cup of milk tea"} shown in \autoref{fig:usage scenarios}(a), the GenAI system operates as an intelligent agent, directly interacting with the user interface to execute a sequence of actions (e.g., navigating apps, entering text, and confirming orders) until the task is fully completed. The scenario in \autoref{fig:usage scenarios}(b) requires the GenAI system to capture, organize, and retain historical user data, enabling it to provide personalized responses based on the user's specific needs.

\textbf{\textit{Concepts.}} To clarify terminology used throughout this paper, we distinguish among several closely related concepts. A \textbf{GenAI smartphon}e refers to the mobile device as a whole, which integrates GenAI capabilities into its operating system and hardware. Within the device, a \textbf{GenAI system} denotes the system-level software stack that orchestrates data access, model invocation, and cross-app actions. The \textbf{GenAI model} is the underlying machine learning model that performs reasoning and content generation; it is the primary entity with which users conceptually interact (e.g., through prompts or natural language instructions).
A \textbf{GenAI agent} represents a task-oriented execution form of the system, in which the GenAI model is empowered to autonomously perceive system context and perform multi-step actions on behalf of the user.
While \textbf{mobile GUI agents} similarly automate user interface interactions, they are typically app-bounded or task-specific and operate with more limited autonomy and system integration. In contrast, GenAI agents in GenAI smartphones are more deeply integrated at the system level, enabling persistent, cross-app behaviors that raise distinct privacy and security considerations.

\subsection{Privacy Challenges in GenAI Smartphones}
\label{privacy implications}
Compared to traditional smartphones and emerging mobile GUI agents, GenAI smartphones have a deeper degree of system integration, autonomy, and persistence, 
giving rise to privacy challenges that extend beyond those of conventional mobile platforms.
For example, GenAI smartphones often require persistent access to sensitive and context-rich data, such as on-screen content, voice inputs, images, and location signals, to support multi-step task execution~\cite{GenAI_phones, wu2024first}. Such access is neither confined to a single app nor limited to discrete interactions, substantially expanding the scope, granularity, and duration of data collection and increasing the risk of over-collection and unintended disclosure.
Beyond direct data access, 
many GenAI features rely on elevated system permissions (mostly accessibility services) to read screen content and perform automated interactions (e.g., clicks and swipes). While enabling agent-like capabilities, these permissions can weaken traditional consent boundaries, introduce risks of permission escalation, and reduce users’ ability to understand and control how their data is accessed~\cite{He2024TheES, Bagdasarian2024AirGapAgentPP, zhang2024privacy}.
Empirical evidence suggests that these risks resonate strongly with users.
A large-scale survey (n $\approx$ 5,000) found that while AI-capable phones offer benefits through on-device intelligence, privacy remains the top concern, with over half of respondents expressing discomfort with AI accessing their personal data~\cite{cybernews2024}. 
Another study revealed that while users often perceive interactions with chatbots as sensitive or highly sensitive, many still disclose health or financial information in practice~\cite{Tran2025UnderstandingPN}, highlighting a persistent tension between convenience and privacy.

While prior work has extensively documented privacy risks in mobile ecosystems~\cite{ntlatywa2024systematic, engel2022mobile, farzand2025systematic, maqolo2025mobile}, the ways in which these GenAI-specific capabilities reshape users’ lived privacy experiences remain insufficiently understood. In this context, investigating users’ privacy perceptions is critical for understanding their concerns and for informing design and policy measures that safeguard data and build trust in this emerging technology.

\begin{figure*}[t]
    \centering
    \includegraphics[width=\linewidth]{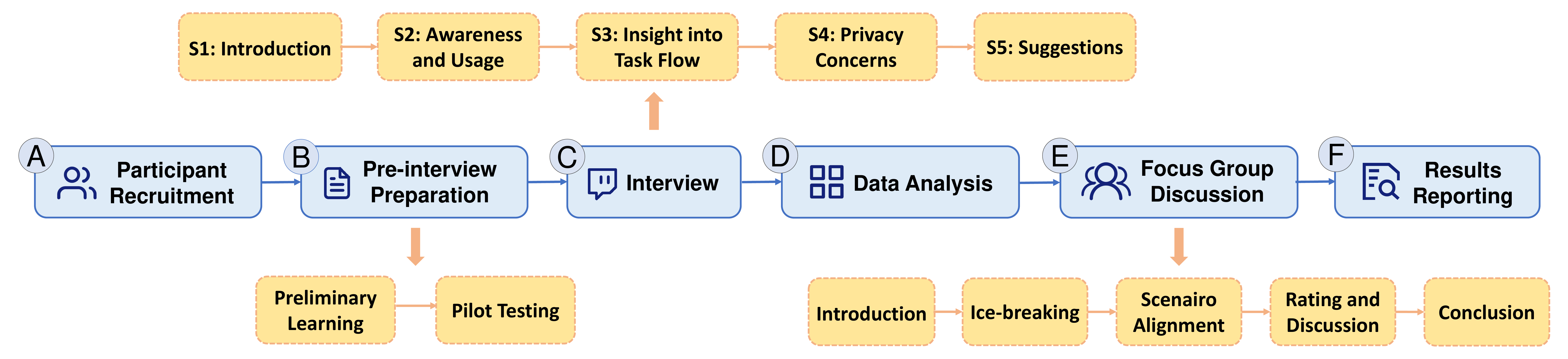}
    \caption{An overview of our study.}
    \label{fig:study flow}
\end{figure*}

\section{Methodology}
\label{sec:methodology}
From June to July 2025, we conducted a semi-structured interview study with 22 participants who are everyday mobile end-users  to explore how they perceive, experience, and respond to the privacy challenges posed by GenAI smartphones. These interviews elicited a rich set of user-identified privacy concerns and design suggestions. 
To complement end users’ perspectives with technical insights, we conducted a follow-up focus group discussion with technically experienced participants to examine the feasibility and implementation implications of their suggestions. 
\autoref{fig:study flow} shows an overview of our study. We next describe each procedure in detail.

\subsection{Participant Recruitment}
\label{sec:recruitment and participants}
Participants were recruited through the author's social media platforms and institutional mailing lists. 
Eligible interviewees were 18 years or older, regular smartphone users, and possessed sufficient familiarity with smartphone usage to meaningfully discuss GenAI-enabled features. 
We did not restrict participation to technical experts; instead, we sought participants with varying levels of technical background, including both individuals working in Information Technology (IT)- or computer science (CS)–related fields and those from non-technical disciplines, provided they had basic awareness of GenAI technologies and GenAI smartphones. 
Our recruitment strategy was motivated by two considerations: 
\one One objective of the study is to elicit user-driven design suggestions for privacy-preserving GenAI smartphones, which presupposes a certain degree of familiarity with GenAI-enabled features and capabilities; and \two including participants without formal technical backgrounds allows us to account for ordinary users’ needs and privacy concerns in real-world settings, thereby reducing expert bias and improving the ecological validity of our findings.
In total, we conducted 22 semi-structured, one-on-one interviews.
We observed thematic saturation after coding the 20th interview, at which point no substantially new concepts or codes emerged; thus, data collection concluded at the 22nd interviews, consistent with established qualitative research practices for theory-building~\cite{doi:10.1177/1525822X05279903}.
Detailed participant demographics are summarized in \autoref{tab:participants}.
Most participants had an academic or professional background in IT/CS fields and reported prior experience with GenAI smartphones. Participants from non-IT/CS backgrounds represented diverse domains, including education, economics, and biology.
All participants held higher education degrees and covered both major mobile ecosystems, with 64\% using Android devices and 36\% using iOS devices.

\subsection{Pre-interview Preparation}
Given that GenAI smartphones are still at an early stage of adoption, some participants had limited prior experience with the technology despite their background in related technologies. 
To facilitate a smoother interview process, we prepared a concise five-page prompting document for participants that introduced key concepts, representative usage scenarios, and examples of privacy management features. The document is approximately 500 words in length and spans 6 pages.
This document was developed by reviewing official materials from major smartphone manufacturers~\cite{Samsung, Apple, Google, HONOR, Xiaomi, OPPO, motorola} and by conducting direct observations of privacy management designs implemented in selected GenAI smartphones. 
Participants were expected to gain a clearer understanding of what constitutes a GenAI smartphone and its potential applications through this prompting document.
This document is available at our repository~\cite{interviews-supplemental-materials}.

Before commencing the formal interviews, we carried out pilot interviews with five individuals from the first author’s research lab. 
During this process, we merged overlapping questions, removed questions not directly related to GenAI smartphones, and made minor adjustments to the order of some questions. 
We also restructured several questions to improve clarity and comprehension. 
To streamline the interview process, we collected participants’ demographic background information in advance via a questionnaire.

\subsection{Interview Process}
\label{sec:interview}
All interviews were conducted by the first two authors.
At the beginning of each interview, participants were informed about the study's purpose, procedure, and data protection measures, after which their consent to audio recording was obtained. 
Half of the interviews were held in person through face-to-face conversations, while the remaining interviews were conducted online via video conferencing to accommodate participants who were unable to attend in person.
Interviews lasted between 40 and 60 minutes, and each participant received compensation of \$10 upon completion.

Our interview protocol consisted of five sections (S1-S5). S1 and S2 introduced key definitions and background while exploring the daily usage habits of participants (RQ1). S3 used concrete examples to illustrate how GenAI smartphones perform tasks, preparing participants for subsequent discussions. S4 focused on eliciting participants’ privacy concerns related to GenAI smartphone usage (RQ2). Finally, S5 gathered participants’ suggestions regarding privacy management and protection based on current smartphone practices (RQ3). 
The complete set of interview questions is available in our repository~\cite{interviews-supplemental-materials}.

\textbf{S1: Introduction.} At the start of the interview, we provided participants with a brief overview of GenAI and GenAI smartphones. 
This introduction included definitions, representative usage scenarios, and key privacy management features.
Much of this content was also included in the pre-distributed prompting document.
The goal of this section was to help participants establish a foundational understanding, ensuring they could comprehend subsequent interview questions without substantial confusion or misinterpretation.

\textbf{S2: Awareness and Usage (RQ1).} For participants currently using GenAI smartphones, we asked their everyday use of GenAI features, focusing on scenarios, frequency, and general usage patterns. 
We also collected all participants' baseline understanding of what a GenAI smartphone is and how it works,  
as well as their initial level of concern regarding privacy on these devices.

\textbf{S3: Understanding Task Execution.} Participants were presented with three interactive, real-world scenarios to illustrate the workflow involved in two representative task types on GenAI smartphones: UI task automation and data processing. 
The purpose was to strengthen participants' understanding of how GenAI smartphones operate during task execution, thereby preparing them for the privacy-related questions in S4 (as some of the questions in S4 will be related to the GenAI smartphone workflow).

\textbf{S4: Privacy Concerns (RQ2).} Participants were asked to discuss their privacy concerns related to data collection on GenAI smartphones. We let participants focus on specific features of GenAI smartphones when performing tasks, such as simulating human interactions and reading screen content.  
By drawing attention to these technical features in detail, we aimed to facilitate participants in reflecting deeply on the potential privacy risks associated with such functionalities and to voice their concerns.

\textbf{S5: Suggestions (RQ3).} Finally, we explored participants' suggestions for protecting personal data and enhancing trust in privacy practices. Guided by the classic Stages of the ``\textit{Data Lifecycle}''~\cite{data_lifecycle}, we designed targeted questions to prompt reflection on suggestions related to data \textit{collection}, \textit{usage and sharing}, \textit{storage}, and \textit{deletion}. 
Additionally, we invited opinions on broader themes such as data transparency, regulatory and policy frameworks, and privacy and security measures designed for vulnerable groups, including the elderly and children.

\subsection{Data Analysis}
\label{sec:analysis}
All interviews were audio-recorded and subsequently transcribed for analysis. During transcription, personally identifiable information (PII) was removed to protect participant anonymity.
We conducted a qualitative analysis using a ``bottom-up'' coding approach, following the principles of thematic analysis~\cite{braun2006using}. 
To ensure rigor, we adopted the team-based coding process outlined by MacQueen et al.~\cite{macqueen1998codebook}, which involves data familiarization, trial coding, codebook development, reconciliation discussions, and consistency checks.
Two authors carried out the data analysis. 
After the formal interviews, each coder thoroughly reviewed the interview transcripts to become familiar with the data, and then independently coded the first three transcripts. Next, their coding outputs were compared, and reconciliation discussions were held to merge similar codes, resolve discrepancies, and develop an initial codebook.  
Using the codes from these three transcripts, we assessed an initial inter-rater reliability (IRR) score with Krippendorff’s alpha~\cite{krippendorff2004reliability,landis1977measurement, marzi2024kalpha, labelstudio2025kalpha}, which was 0.59. 
The coders further independently coded another three transcript, the scores reached 0.82, indicating a substantial level of agreement. 
This process aimed to evaluate the practical consistency of the preliminary codebook and ensure the reliability of subsequent divided coding~\cite{mcdonald2019reliability}. 
Next, both coders divided the rest of the transcripts for coding and engaged in regular discussions to iteratively refine the codebook. The final codebook included 10 themes, 24 subthemes, and 95 codes, available in our repository~\cite{interviews-supplemental-materials}.

\subsection{Focus Group Discussion}

Following the interviews, we collected a broad range of user-generated suggestions for enhancing privacy in GenAI smartphones (RQ3). However, given the varying levels of technical expertise among interview participants, these suggestions differed in their specificity, technical grounding, and practicality. To further examine how such user-driven ideas could be realized in practice, we conducted a focus group discussion to assess the proposed suggestions from a technical perspective.

The focus group method was chosen because it facilitates structured and interactive discussion, enabling collective reflection and the integration of complementary technical perspectives more effectively than isolated individual assessments~\cite{morgan1997focus, kitzinger1994methodology, o2018use}. Based on the interview findings, we carefully selected seven representative suggestions (\autoref{tab:privacy_recommendations}) to guide the discussion.
The session was moderated by the first two authors and involved eight participants with strong technical backgrounds in GenAI and mobile systems, all of whom had prior experience with GenAI smartphones and had also participated in the earlier interviews. Their expertise enabled a thorough discussion from feasibility and implementation perspectives.
At the start of the session, the moderators introduced the study goals and procedure. The discussion then proceeded through the seven distilled suggestions.
For each suggestion, participants first provided individual ratings using a standardized five-point Likert scale along three dimensions: \one \textbf{effectiveness} in mitigating privacy risks; \two \textbf{feasibility} within existing technological, regulatory, and product constraints; and \three technical implementation \textbf{difficulty} and associated engineering costs.  
These quantitative ratings were followed by open-ended group discussions, during which participants elaborated on their considerations around each dimension.
The focus group lasted approximately 60 minutes and was fully audio-recorded, with key points documented by the research team. 
All collected data were used exclusively for this research and were anonymized during subsequent analysis. The discussion protocol is available in our repository~\cite{interviews-supplemental-materials}.

\subsection{Limitations}
\label{sec:limitations}
As with many qualitative studies, our interview-based research is subject to the limitations of self-reported data.
Participants’ responses may be influenced by recall bias, leading to incomplete or inaccurate accounts of their experiences.
The framework of interview studies can potentially limit participants’ responses. 
To reduce this impact, we minimized interviewer guidance and ensured that questions remained neutral, offering prompts only when participants were unable to generate a response. 

Additionally, our study has limitations related to participant selection. While the sample size (n = 22) was sufficient for data saturation, it limits the generalizability of the findings. Recruitment  
primarily focused on younger users and individuals with relevant technical expertise, as they are more likely to engage with GenAI smartphone features and share reflections on emerging privacy implications, making them particularly informative for this exploratory study.
Nevertheless, this focus may underrepresent other user groups. Future work could extend this research by engaging a more diverse and representative user population.

Finally, since GenAI smartphones are still an emerging technology, not all participants had extensive hands-on usage experience, which may have influenced their understanding of certain technical features and privacy risks. 
To support meaningful discussion, we provided a preliminary learning document to participants outlining GenAI smartphone features and privacy-related settings, and the interview questions explicitly focused on security and privacy issues. 
While this helped establish a shared baseline for the interview, it may also have shaped participants’ attention toward potential risks.
We view this as a necessary trade-off for studying an emerging technology and encourage future work to complement our findings with longitudinal or in-situ studies as GenAI smartphones become more widely adopted.

\begin{figure*}[t]
    \centering
    \includegraphics[width=\linewidth]{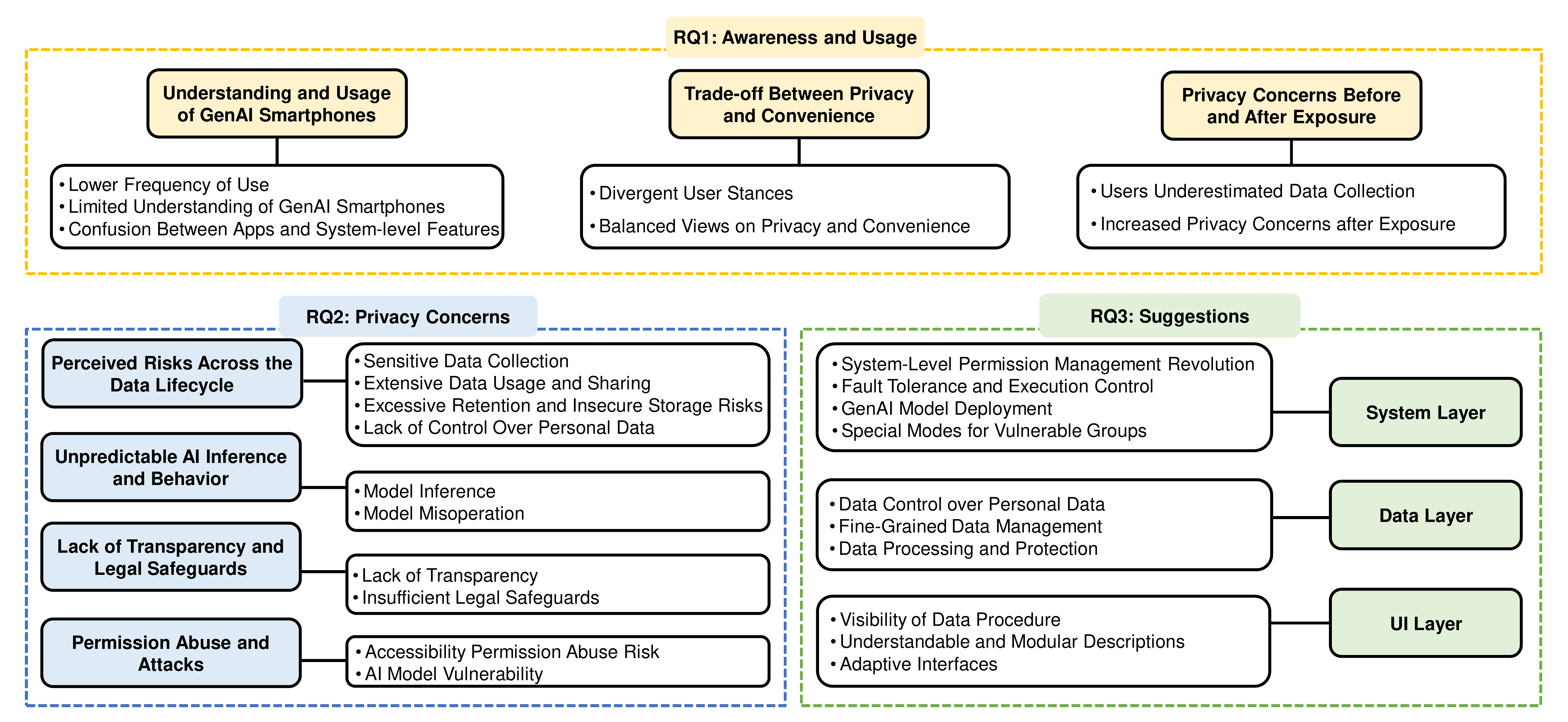}
    \caption{The findings summarized from the interview.}
    \label{fig:Findings Summary}
\end{figure*}

\section{RQ1: Awareness and Usage}
\label{awareness and usage patterns}
Before examining specific privacy concerns, we first explored participants' baseline understanding of GenAI smartphones, their usage patterns, general attitudes toward privacy-convenience trade-offs, and privacy concerns before and after understanding system mechanisms. These foundational insights help contextualize the more detailed privacy concerns discussed in subsequent sections. Key findings are summarized in the yellow region of \autoref{fig:Findings Summary}.

\subsection{Understanding and Usage}
Understanding users' knowledge and behavioral patterns is crucial for assessing adoption and identifying risks in emerging technologies. Prior work, such as the \textit{Technology Acceptance Model}~\cite{davis1993user}, suggests that users' understanding significantly influences their willingness to adopt and actual usage behavior. 

Although participants generally reported having some familiarity with GenAI smartphones, we observe a clear gap between their \textit{perceived understanding} and their \textit{actual knowledge and use} of system-level GenAI capabilities.
Many participants tended to conceptualize GenAI smartphones at the app level, assuming that simply having GenAI apps installed qualifies a device as a GenAI smartphone. Even among those who recognized that such devices integrate GenAI models at the system level, most were unaware of underlying mechanisms such as the use of accessibility permissions to access on-screen content and perform automated interactions. 
This knowledge gap has direct implications for privacy awareness: when users do not understand how GenAI features access and process data, they may underestimate privacy risks, grant permissions without fully understanding their scope, or fail to adopt appropriate protective measures. 

For participants currently using GenAI smartphones, we further examined the frequency of their use of built-in GenAI features (excluding standalone GenAI apps). In terms of self-reported usage, only one participant (P22), a researcher working on mobile agents, reported daily use of GenAI-related features. The most (13 out of 15 participants) engaged with GenAI functionalities only occasionally, primarily for relatively simple tasks such as interacting with intelligent assistants, photo editing, or screen translation. To understand this limited adoption, we drew on \textit{Rogers' Diffusion of Innovation Theory}~\cite{rogers2003diffusion}, 
and our findings highlight two key factors shaping usage: \textit{perceived advantage} and \textit{observability}.
First, four participants perceived limited advantages of GenAI smartphones over traditional devices. As P6 noted, \textit{``I don't think it's smart enough yet to make a big difference in my life.''} Concerns about accuracy in complex tasks further reduced users’ willingness to rely on GenAI features, leading many to prefer manual operation. 
Second, the observability of GenAI smartphones' distinctive capabilities remains low, leaving users unclear about what these devices can actually do. As P11, lacking a CS/IT background, stated, \textit{``I didn't realize GenAI smartphones could perform such advanced tasks now.''} Even technically proficient users like P14 admitted, \textit{``Honestly, I'm unclear about which features utilize LLMs, so I feel vague about GenAI smartphone capabilities.''} This pattern suggests users have not yet discovered the full capabilities of GenAI smartphones, constraining their usage to familiar and low-risk tasks rather than exploring advanced GenAI-driven features.

\subsection{Trade-off Between Privacy and Convenience}
Given that GenAI smartphones promise enhanced convenience through automated task execution while requiring extensive personal data access, understanding how users navigate the privacy-convenience trade-off becomes critical for designing systems that align with user values.
We explored users’ perspectives on the trade-off between privacy and convenience, categorizing them into three roughly equal groups: those who prioritized convenience (n=7), those who remained neutral (n=8), and those who prioritized privacy (n=7). 
Participants who prioritized convenience tended to use GenAI smartphones to ease their life. 
As P21 noted: \textit{``In fact, our data are of little use when kept in our own hands. It only becomes valuable when shared accessibly and reasonably utilized.''}
However, these users did not disregard privacy entirely; rather, they placed the responsibility on smartphone manufacturers to manage and protect their personal data appropriately.
In contrast, participants who prioritized privacy were more concerned about the risks of automated massive data collection. 
When using GenAI smartphones, they preferred manual operation of devices over AI-enabled automation. 
Overall, participants’ views on privacy and convenience were divided yet balanced, underscoring that privacy considerations must remain central in the design of GenAI smartphones, even as users seek greater convenience.

\subsection{Privacy Concerns Before and After}
We measured participants’ privacy concerns using a five-point scale, where one indicated no concern at all and five indicated extreme concerns. 
Ratings were collected both before and after participants completed S3 and S4. 
Compared to their initial rating, 41\% of participants reported increased concerns in the second assessment, typically by one or two levels, their average level of privacy concern rising from 2.9 to 4.1.
These findings suggest that, prior to understanding the inner workings of GenAI smartphones, many users greatly \textbf{underestimated} the scope of data collection and demonstrated a relatively low level of privacy awareness. 
As P2 noted, ``\textit{I hadn't thought about these issues before.}'' 
After being exposed to the implementation details, participants expressed greater concern about privacy risks.
21 participants rated their concerns as neutral or higher, with only one participant remaining unconcerned, indicating that most users still maintain a high level of privacy concern and remain vigilant about the potential privacy risks of GenAI smartphones.
This shift underscores the importance of \textbf{transparency} in raising privacy awareness and highlights the need for manufacturers to provide clear and authentic privacy notices. 
For GenAI smartphones, this means disclosure must extend beyond traditional app permissions to explain cross-app data flows, automated decision-making processes, and the full scope of GenAI models access, enabling users to make truly informed choices about their privacy.

\section{RQ2: Privacy Concerns}
\label{privacy concerns}
After analyzing users' basic awareness and usage patterns, we gained deeper insights into their specific privacy concerns. The following sections explore perceived risks across the data lifecycle, unpredictable AI inference and behaviors, lack of transparency and legal safeguards, and permission abuse and attacks. Key findings are summarized in the blue part of \autoref{fig:Findings Summary}.
\subsection{Perceived Risks Across the Data Lifecycle}
Users reported various privacy risks at each stage of the data lifecycle, from collection to processing, storage, sharing, and deletion.

\textit{\textbf{Data Collection.}} During the data collection phase, users' privacy concerns arise from the sensitive nature of the information collected, which can be categorized into active provision and passive collection. 
Regarding active data provision, users intentionally upload information to facilitate interactions, but such inputs may inadvertently include highly sensitive content. 
For instance, some users upload work-related materials to enhance productivity. 
As P16 noted: \textit{``There are indeed many tasks that are more convenient to handle with AI, so I have already uploaded a lot of things that shouldn't have been uploaded to AI.''} 
Regarding passive data collection, users' concerns primarily stem from their lack of effective control over the data collection. 
The lack of sufficiently granular data permission management mechanisms prevents users from effectively regulating what data their devices collect. 
As P14 expressed: \textit{``AI systems have relatively higher permissions. They do not only collect user information,  but also information related to third-party apps, which is a very dangerous thing.''}
The possibility that GenAI smartphones may be continuously running in the background and collecting user data through accessibility permissions could further fuel concerns. 
As P1 stated: \textit{``If it is constantly reading my screen in the background, I think that's very risky, because I cannot avoid clicking on my personal information, ID card, bank card, and other private details.''} 
These concerns reflect users' fundamental uncertainty about when, what, and how much data is being collected during their daily GenAI smartphone use unconsciously.

\textit{\textbf{Data Usage and Sharing.}} During the data usage and sharing phase, participants were concerned that personal information could be exploited without consent or a clear purpose. 
They feared that manufacturers might use extensive personal data to build detailed profiles–including health conditions, social networks, behavioral habits and communication styles–for targeted advertising or share it with third parties. As P11 noted, \textit{``GenAI can infer which diseases you have based on health data, which could lead to employment discrimination if your employer knows this. Alternatively, it can reconstruct your social network based on chat messages and notifications, data that is crucial for scammers.''}
Participants also noted that data leaks could lead to fraud, causing both financial and emotional losses. 
Location tracking was identified as a particularly serious concern. 
As P2 stated: \textit{``... Location information involves personal safety, should be strictly protected because it can pinpoint an individual's location, and I think that's very dangerous... I would feel completely under the control of the smartphone manufacturer, which would make me feel very unsafe.''} 
The safety risks of location tracking become especially acute when such data is shared with third parties or leaked to malicious actors. Once location information is combined with other behavioral signals, attackers can not only track where users are, but also predict when they are most vulnerable.
P9 further elaborated on safety concerns: \textit{``... If data obtained from my smart wearable device is leaked, someone who wants to interfere with your life could easily use this data to determine when you are sleeping and carry out attacks or dangerous actions during that time.''}
Participants also expressed concerns about insecure transmission processes, such as the lack of encryption measures, which could lead to data interception and misuse during data sharing. 

\textit{\textbf{Data Storage.}}
During the data storage phase, participants primarily expressed concerns about excessive data retention and security vulnerabilities. 
They pointed out that retaining data longer than necessary enlarges the attack surface and increases exposure risks. As P2 stated: \textit{``Excessive storage of user data exacerbates the risk of privacy breaches.''} 
Additionally, participants were concerned about data stored in the cloud or accessed through APIs.
As P3 stated: \textit{``Data is stored in the cloud, so the first question is whether the vendor's own security management is strong enough to withstand hackers stealing data.''}
They worried that inadequate safeguards could increase the risk of data breaches and identity theft. 

\textit{\textbf{Data Update and Deletion.}}
Participants emphasized the importance of maintaining control over personal data. Once data is collected, users often feel powerless to delete, modify, or manage the information stored about them. 
As P4 stated: \textit{``Can I update or delete it? I don't even know; I thought this information was stored iteratively.''} Many participants thought they lacked effective means to exercise such rights, which reinforces their perception of vulnerability.

\subsection{Unpredictable AI Inference and Behaviors}
Participants raised concerns about privacy and safety threats inherent in GenAI models’ reasoning and action mechanisms.

\textbf{\textit{Model Inference.}} 
A primary concern is that models may infer additional PII from the data they collected. 
Prior research has shown that GenAI models could derive sensitive attributes that individuals may not wish to disclose~\cite{Neel2023PrivacyII, Tmeke2024PrivateAI, Staab2023BeyondMV}, such as users' political views. 
As P22 stated: \textit{``GenAI may infer other sensitive information based on collected personal data. 
For example, if you frequently order takeout from a specific address, the model might infer this is your residence and then push ads for nearby services.''}
Additionally, participants mentioned that GenAI's reasoning capabilities extend to linking already collected information to data available online. 
As P3 mentioned: 
\textit{``... if you don't want the model to know your phone number, but the model can infer it from other information combined with phone numbers stored in the training data, it can link your private information without authorization.''}

\textbf{\textit{Model Misoperation.}} 
When performing multi-step tasks, GenAI smartphones rely on GenAI models to reason and generate execution steps. 
However, participants expressed concerns that models may misunderstand user intent or generate incorrect steps due to phenomena such as hallucinations. 
P14 stated: \textit{``If, as you mentioned, the phone uses an LLM to generate a click operation, and if the LLM makes an incorrect judgment and uses accessibility permissions to click on an unintended location, I believe this poses a risk.''} 
The consequences range from exposing sensitive information to unintended recipients to authorizing financial transactions with password-free payment enabled. P10 illustrated several concerning scenarios: \textit{``It might accidentally delete something important, or when ordering bubble tea, accidentally order multiple cups and complete the payment. It could also swipe to other pages and access content I didn't intend to show it—basically obtaining more information than I wanted to give.''} 

\subsection{Lack of Transparency and Legal Safeguards} 
Beyond the technical risks, participants expressed concerns regarding the non-transparency of GenAI smartphone operations and the inadequacy of regulatory protections.

\textbf{\textit{Lack of Transparency.}}
Participants highlighted the lack of transparency in data processing by GenAI smartphones, including the underlying reasoning process and data flows which is useful for users to gain insight into the task execution workflow. 
The reasoning chain supporting task completion is hidden, and the sequence of steps is not provided to users for inspection or correction. This opacity of task execution raised concerns about potential misuse. Some participants worried that manufacturers may intentionally use unnecessary operations to obtain user information. As P6 mentioned: \textit{``I think smartphone manufacturers may intentionally make mistakes or include unnecessary steps to obtain some of your information through these extra steps.''} 
P1 similarly remarked, \textit{``It's like handing your phone to a stranger. I have no idea what he is doing with it.''} This lack of visibility into data processing leaves users unable to assess whether the system is collecting only the information necessary for the requested task or gathering additional data without justification. 

\textbf{\textit{Insufficient Legal Safeguards.}}
Participants highlighted the lack of a robust legal framework to govern data collection and use in GenAI smartphone ecosystem. 
They noted a fundamental power imbalance between users and manufacturers, a phenomenon scholars describe as \textit{``digital resignation''}–the reluctant acceptance of privacy intrusions due to perceived lack of alternatives or agency~\cite{draper2019digital}. 
As P6 stated: \textit{``Now the phone manufacturers are also entering the fray to collect my data. \ul{I am using their phone now, and if they want to collect my data, what can I do?} I have no way to fight back...I just hope that there are some macro-level, legal constraints for these phone manufacturers.''}
Overall, participants' concerns originated in the inability to effectively constrain GenAI smartphone behavior without stronger legal protections.

\subsection{Permission Abuse and Attacks}
Participants also identified specific security vulnerabilities that could be exploited to compromise user data and device control in GenAI smartphones.

\textbf{\textit{Accessibility Permission Abuse.}} 
As aforementioned, GenAI smartphones often rely on elevated system-level permissions, e.g., accessibility services, to read screen content and perform automated actions.
Participants expressed strong concerns that once granted, these permissions provide extensive and persistent access to system-level data and user activities, making them susceptible to malicious exploitation.
For example, P2 shared a friend's experience with telecom fraud, in which attackers gained continuous remote control of the phone. This incident heightened P2's concern that accessibility permissions could be abused to manipulate his own device, leading to both financial and emotional harm. 

\textbf{\textit{AI Model Vulnerability.}} Participants acknowledged that GenAI models themselves are vulnerable to attack or manipulation, and worried that such risks may be amplified in smartphone environments. They identified two main categories of threats: operational process attacks and malicious data acquisition. 
First, participants expressed concerns about adversarial inputs or poisoned data that could mislead models into generating harmful actions when embedded in automated workflows. As P22 stated: \textit{``I think that some attacks on LLM, such as adversarial attacks and backdoor attacks, may cause the model to go beyond the scope of instructions and do some malicious things.''}
Second, participants noted that prompt injection or model inversion attacks might enable attackers to extract personal information processed by the model or infer sensitive data from prior interactions. 
As P13 explained: \textit{``It seems that there are potential inference attacks targeting GenAI data. 
If a phone uses my data for training, it is possible for an attacker to infer the data used by the GenAI for training. In that case, the GenAI could potentially leak my data.''}

\section{RQ3: Suggestions}
\label{sec:suggestions}
Building on the privacy concerns identified above, we further explore potential approaches to mitigating these risks by eliciting design suggestions from participants’ perspectives.
After the interview, we coded these user-generated suggestions into a three-layer framework, i.e., system, data, and UI layers (\S\ref{sec:Interview suggestions}). 
We then further examined and reflected on some suggestions through a focus group discussion with tech-savvy participants (\S\ref{sec:fg-suggestions}).

\subsection{Suggestions Derived from the Interview}
\label{sec:Interview suggestions}
This section provides a detailed summary of user suggestions raised during the interviews, as shown in the green part of \autoref{fig:Findings Summary}.

\subsubsection{System Layer}
The system layer focuses on privacy protections embedded at the OS level, where GenAI models are granted access to device resources and endowed with autonomous execution capabilities. Suggestions in this layer address foundational mechanisms that govern how GenAI operates across the system.

\textbf{\textit{System-Level Permission Management Revolution.}} 
Regarding permission management, participants proposed solutions ranging from incremental refinements within the existing framework to more fundamental architectural changes.
As an incremental improvement, participants emphasized the need for explicit re-authorization when GenAI systems attempt to access or act upon third-party apps. As P13 explained: \textit{``
Actually, when users first authorize an app, they may not have carefully read it at all. 
I think it's still necessary for GenAI to reinform users about the data being collected and the permissions that may be used.''}
Moreover, they emphasized the need to \textit{separate permission management for GenAI systems and third-party apps}, rather than allowing them to use system-level accessibility permissions. 
Besides, several participants argued that such refinements are ultimately insufficient. They pointed out that current permission frameworks 
(originally designed for app-bounded, user-initiated interactions) might no longer be adequate for the emerging management demands introduced by system-level GenAI integration.
P15 emphasized, \textit{``Applying the original permission management mechanism to AI agents is highly unreasonable, as it essentially constitutes an operation that exceeds authorized permissions.''} 
As a result, participants called for fundamental, system-level redesigns of permission management that explicitly account for the characteristics of GenAI.
As P2 proposed: \textit{``To comprehensively solve this problem, it should be done at the mobile system level, optimizing the existing permission management mechanism and making adjustments based on the characteristics of GenAI.''} 

\textbf{\textit{Fault Tolerance and Execution Control.}}
Fault tolerance in GenAI systems refers to its robustness to component faults during sensitive or multi-step tasks. 
In GenAI smartphones, a key strategy is to \textit{automatically terminate execution and alert the user to take over} when a failure or unexpected state occurs. 
For example, if a payment transaction fails, the system should stop immediately rather than attempt a risky continuation. 
Furthermore, the system should issue a clear fault notification, explaining the problem and requesting user confirmation before continuing. 
P13 said: \textit{``When it makes a mistake, I hope \ul{I can take over my phone immediately}.''}
Some participants suggested restricting the operational scope of GenAI based on different task scenarios. 
As P10 said: \textit{``Based on the task, determining the scope and setting a range within which the GenAI can only operate. If it exceeds this range, stop or remind the user.''} For example, during a navigation task, the system could be prevented from opening social apps. Participants viewed this approach as helping prevent accidental operations or unintended app launches at the source.

\textbf{\textit{GenAI Model Deployment.}}
Participants emphasized that the deployment strategy directly affects the volume of personal data exposed, and therefore the overall trustworthiness of the GenAI systems. 
Participants stressed that \textit{local deployment of models should be prioritized whenever feasible}. 
They noted that running GenAI models directly on the device reduces the need for continuous data transmission to cloud servers. 
As P6 firmly stated: \textit{``\ul{I only accept large models deployed locally.} Even if not uploading to the cloud might sacrifice some capabilities of the large models.
But if it involves aspects like internet connection or uploading to the cloud, I hope I can be given the option to choose whether to accept it each time.''}

\textbf{\textit{Special Modes for Vulnerable Groups.}}
Participants emphasized that vulnerable groups, such as the elderly and children, face heightened privacy and safety risks when interacting with GenAI smartphones.
For this, participants advocated for system-level mechanisms that combine assisted privacy management with stricter control over sensitive permissions. 
As P19 stated, \textit{``I believe it is necessary to implement child-friendly and senior-friendly modes. For children, certain features may need to be blocked, especially those involving financial transactions. For seniors, the focus should likely be on incorporating anti-phishing measures or similar functionalities.''} 
Participants suggested that, by default,  high-risk operations such as password-free payments, large-value financial transactions, or direct system-level modifications should be disabled or gated behind additional safeguards. 
In addition, participants proposed incorporating guardian-assisted workflows into the permission and execution model. 
As P14 said, \textit{``In certain extreme situations, guardians may intervene in to assist the elderly or children to operate certain tasks.''} 
In those scenarios, when the special mode is turned on, the system could require remote confirmation or intervention from a guardian. 
Participants perceived this cooperative mechanism as essential.

\subsubsection{Data Layer}
The data layer governs the information lifecycle for privacy protection by defining how user data is collected, processed, and managed. Suggestions in this layer focus on strengthening user control over personal data, fine-grained data management, and enhancing data processing and protection mechanisms.

\textbf{\textit{Data Control over Personal Data.}}
Participants emphasized the importance of giving users explicit control over the data collected and processed by GenAI smartphones. 
They expressed a desire to review, correct, and delete personal data, as well as decide how such data may be used afterward. 
P17 suggested: \textit{``I think it is necessary to design a dedicated data page for GenAI usage, where users can update and delete the data collected by GenAI.''} 
This reflects an expectation for a centralized data management space that provides transparent oversight and control over GenAI's data usage.

\textbf{\textit{Fine-Grained Data Management.}}
Of all the participants, 17  consistently emphasized the need for fine-grained data management mechanisms that give users precise control over what data the smartphone can access, collect, and retain. 
They noted that current GenAI smartphones largely inherit traditional ``all-or-nothing'' data access models, forcing users to either accept broad data access to use it or forgo the smartphone entirely.
Users are expected to exercise \textit{fine-grained control over the data} that GenAI can access. 
As P1 stated: \textit{``I may only want it to process a certain image, but it actually has the permission to access my other images or files.''} 
Several participants proposed shifting from function-oriented controls to information-oriented data management. P3 explained: \textit{``In addition to this function-oriented design, could you consider making it information-oriented? For example, I could explicitly specify which types of information I do not want the assistant to access, such as location, contacts, ID, etc., rather than achieving this by restricting a specific function.''} 
Such an approach could reduce the configuration burden associated with managing data access on a per-feature basis, enabling users to implement fine-grained data management more easily and scalable as GenAI capabilities expand.

\textbf{\textit{Data Processing and Protection.}}
Participants emphasized that essential security measures of personal data by smartphone manufacturers are critical safeguards against privacy breaches. As P14 said: \textit{``I think data protection is the most important thing. Whether the user is aware of/authorizes it or not, if there are problems with their data protection and it causes privacy leaks, everything else becomes meaningless.''}
They identified three essential measures: \textbf{data pre-screening}, \textbf{anonymization}, and \textbf{encryption}. 
Data pre-screening was highlighted as a particularly important feature that they expect smartphone manufacturers to implement. As P3 noted: \textit{``The system could provide a list of information, a data summary, or a report detailing potentially collected private data, enabling users to select which items they do not wish to be uploaded or used.''} 
Participants proposed that the system could automatically detect potentially sensitive elements (e.g., names or personal identifiers) and flag them for the user, prior to any data upload or storage, so that users can determine which data can be shared with AI services and which must strictly remain private.
They also emphasized the importance of anonymization prior to storage or processing and encryption must be consistently applied during both storage and transmission.

\subsubsection{User Interface Layer}
The UI layer enhances user transparency and awareness by designing user-friendly interfaces for privacy-related actions. Suggestions for this layer focus on making GenAI data practices visible, understandable, and actionable for users.

\textbf{\textit{Visibility of Data Procedure.}}
Participants stressed the importance of making GenAI data processing practices visible, clear, and understandable. 
They expected manufacturers to explicitly inform them about what data GenAI collects, whether it is uploaded to the cloud, and whether encryption and anonymization are applied.  
As P1 stated, \textit{``I hope they will provide me with clear step-by-step instructions on what they did, that is, the operating steps, what information they uploaded to GenAI, and then what information it downloaded.''} 
Additionally, some participants who had experience with reasoning LLMs (e.g., DeepSeek-R1~\cite{deepseek}) expressed that they would like GenAI smartphones to make the reasoning process explicitly visible during task execution, so that users can monitor ongoing actions and the data being accessed.
Participants also noted that although manufacturers have drafted privacy policies for GenAI systems, these documents are often lengthy and difficult to read.
They advocated for more usable context-aware notifications, such as pop-ups or system alerts, when GenAI accesses sensitive data or performs high-risk operations (e.g., payments or cross-app interactions). 
As P8 stated: \textit{``Privacy-related disclosure should be presented in a more prominent way, like a pop-up window; 
they should immediately inform you what to do when you encounter a problem.''}
Such just-in-time disclosures help users make informed decisions at the moment their data is actually being used.

\textbf{\textit{Understandable and Modular Descriptions.}} 
Participants emphasized that privacy communication should be understandable and modular, so that users can quickly grasp what data is accessed, why, when, and by whom.
To do this, participants suggested breaking down GenAI permissions and data practices into small, inspectable “modules” and presenting them using plain language. For example, each module could \one name the data type (e.g., screen text, photos, location), \two specify the operation (read/upload/retain/share), \three state the purpose tied to a task, and \four indicate scope and duration (one-time vs. continuous; selected items vs. all). They also advocated avoiding ``one-switch-for-many'' designs by exposing per-module toggles (or expandable cards) so users can review and adjust specific accesses without disabling an entire GenAI feature.

\textbf{\textit{Adaptive Interfaces.}} 
Participants emphasized that adaptive interface design should reflect users’ differing privacy preferences and standards, rather than adopting a one-size-fits-all approach.
They noted that users vary widely in their tolerance for data access, desired level of control, and ability to process privacy information, and therefore may require different interface configurations and feature sets.
Participants suggested that privacy feedback and controls should be adaptively embedded into the interaction flow, with the level of detail, frequency of prompts, and form of presentation adjusted based on users’ privacy sensitivity and needs.
For example, P18 suggested that systems should flag sensitive content and request user confirmation before further data collection: \textit{``If it needs to collect our images, it could first locally identify and flag potentially sensitive images, and then prompt the user. Only after receiving user confirmation should it proceed with further collection.''}  
Participants further noted that such adaptive presentations are particularly important for elderly users and children, for whom simplified language, clearer explanations, and multimodal cues can reduce misunderstanding and interaction errors. 

\subsection{The Practicability of Suggestions}
\label{sec:fg-suggestions}

%
%
\begin{table*}[t]\small
\centering
\caption{Selected privacy design suggestions for GenAI smartphones discussed in the focus group. The last three columns present the distribution of ratings (1-5 from left to right) assigned by eight participants, with ``mean ± standard deviation''. Notably, we adjust the difficulty scale, making higher values indicate that the suggestion is technically easier to implement.}
\label{tab:privacy_recommendations}
\begin{tabularx}{\textwidth}{lXScScSc}  
\toprule
\textbf{Layer} & \textbf{Recommendation} & \textbf{Effectiveness} & \textbf{Feasibility} & \textbf{Difficulty} \\
\hline
\multirow{8}{*}{System Layer}
& \textbf{Permission Isolation}: Separate permission management for GenAI systems and third-party apps instead of sharing a unified accessibility permission. 
 & \makecell{\includegraphics[width=0.8cm]{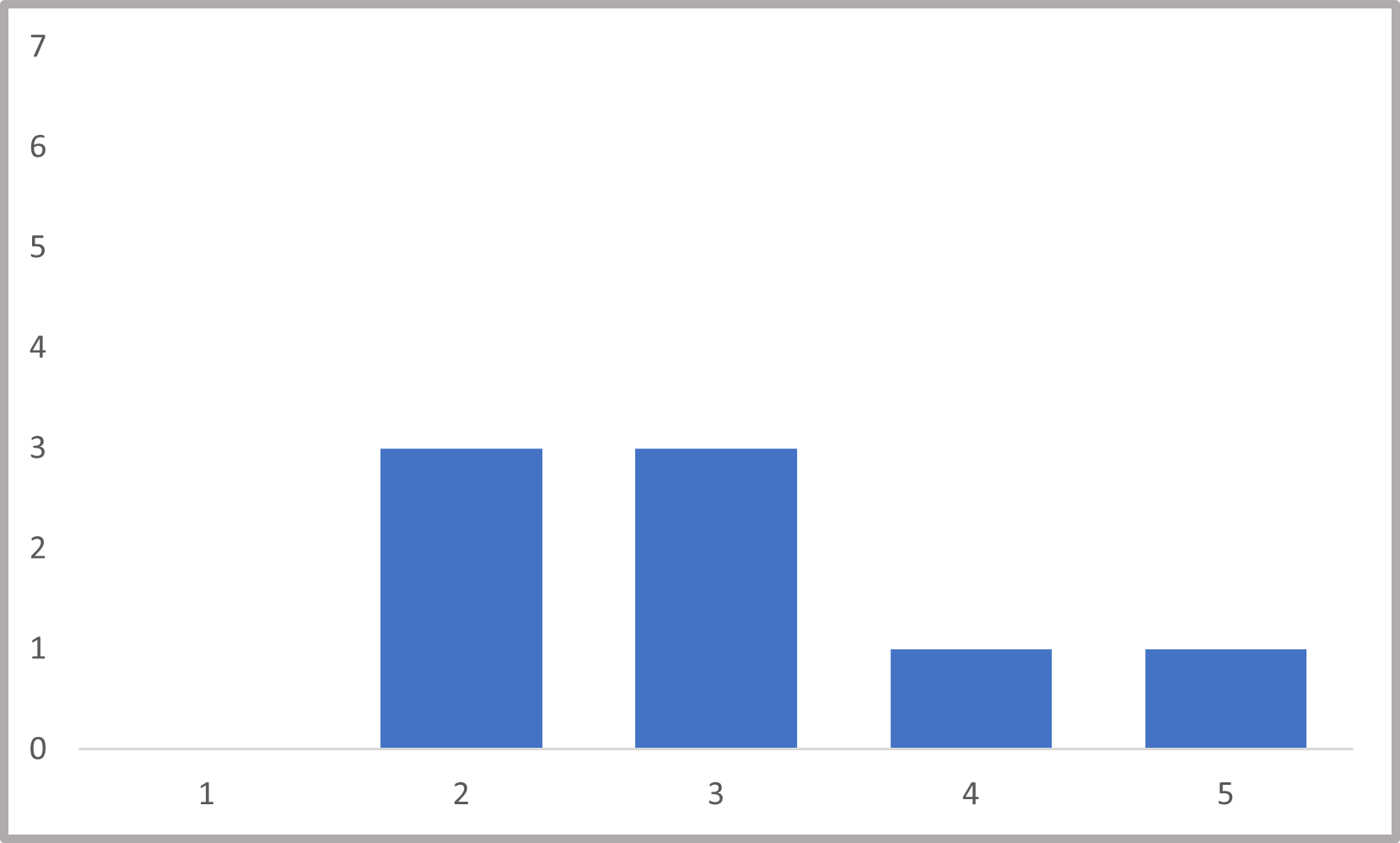}\\[-1pt]{3.0\tiny±1.1}}
 & \makecell{\includegraphics[width=0.8cm]{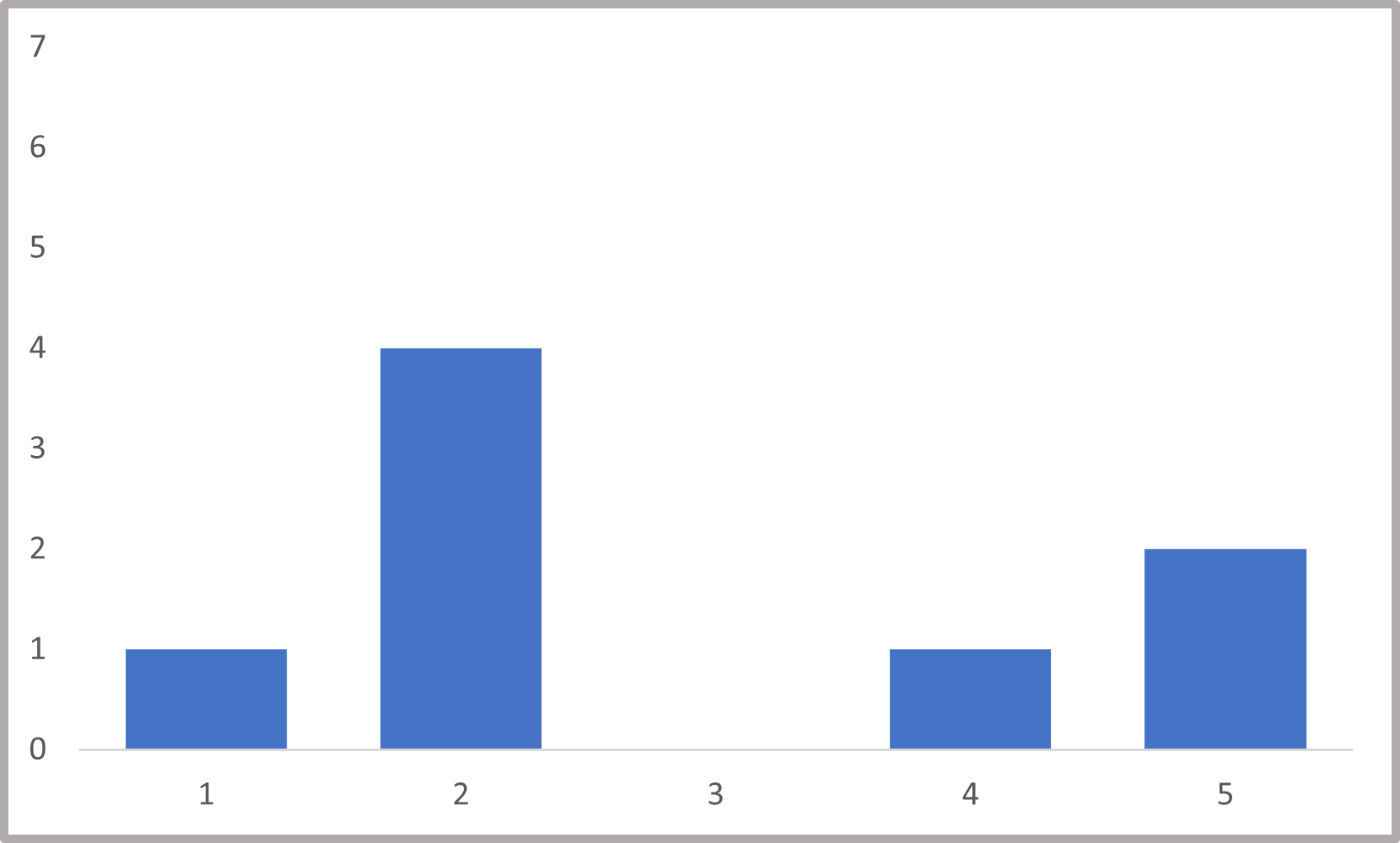}\\[-1pt]{2.9\tiny±1.6}}
 & \makecell{\includegraphics[width=0.8cm]{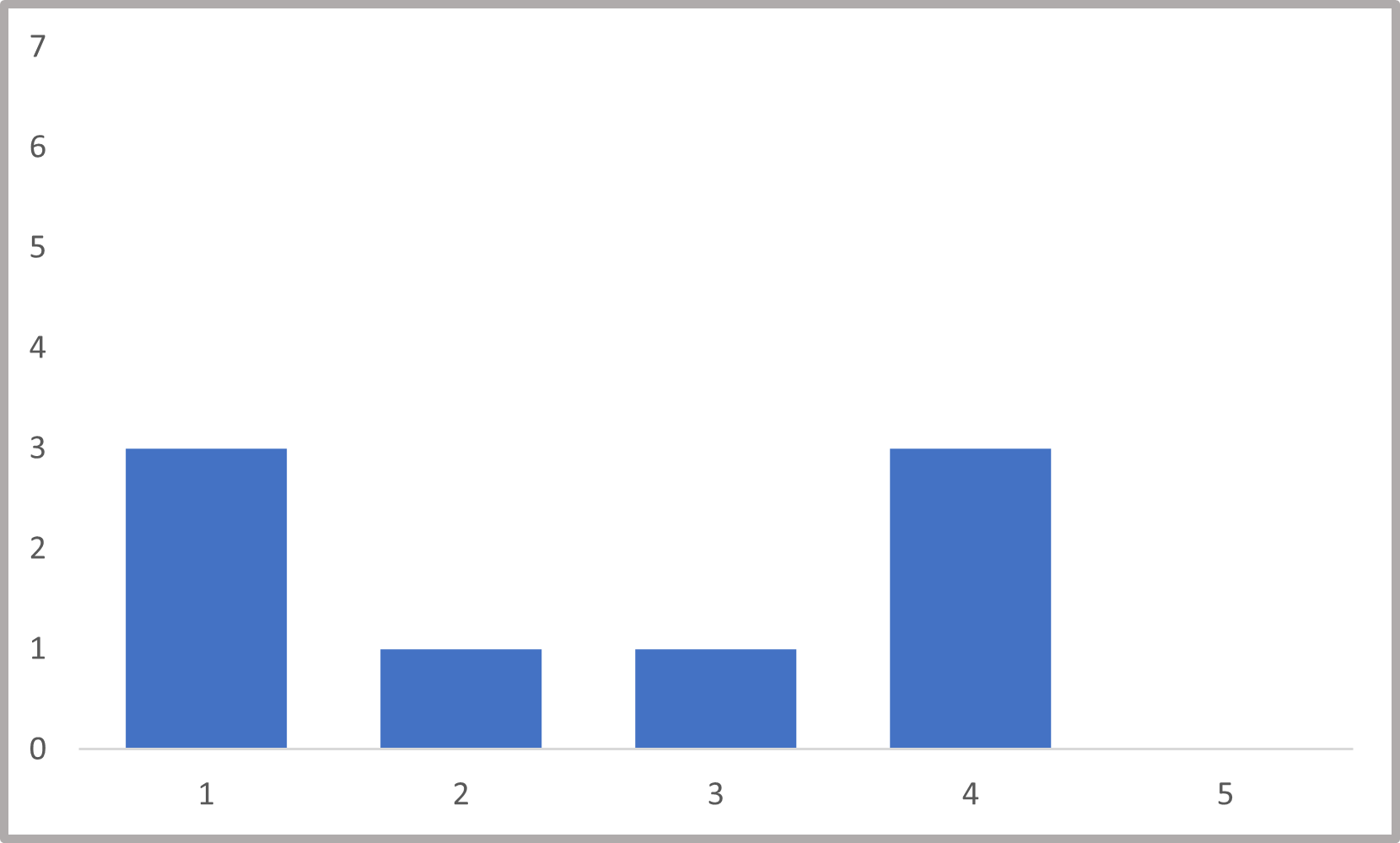}\\[-1pt]{2.5\tiny±1.4}} \\
\cline{2-5}
& \textbf{Intent-Based Permission Framework}: Shift from traditional app-centric permission models to an intent-based framework specifically designed for GenAI systems. 
 & \makecell{\includegraphics[width=0.8cm]{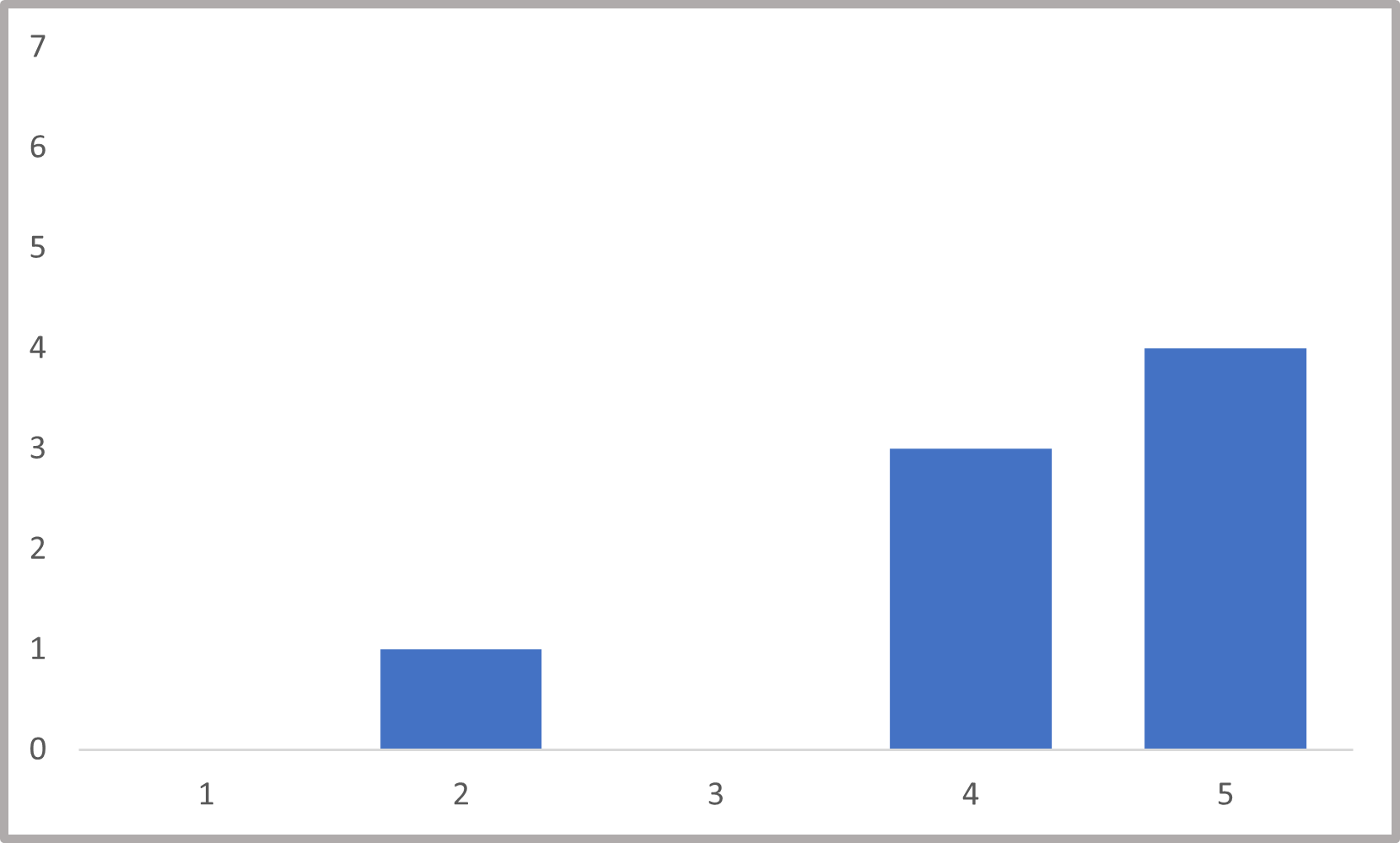}\\[-1pt]{\textcolor{green2}{4.3}\tiny±1.0}}
 & \makecell{\includegraphics[width=0.8cm]{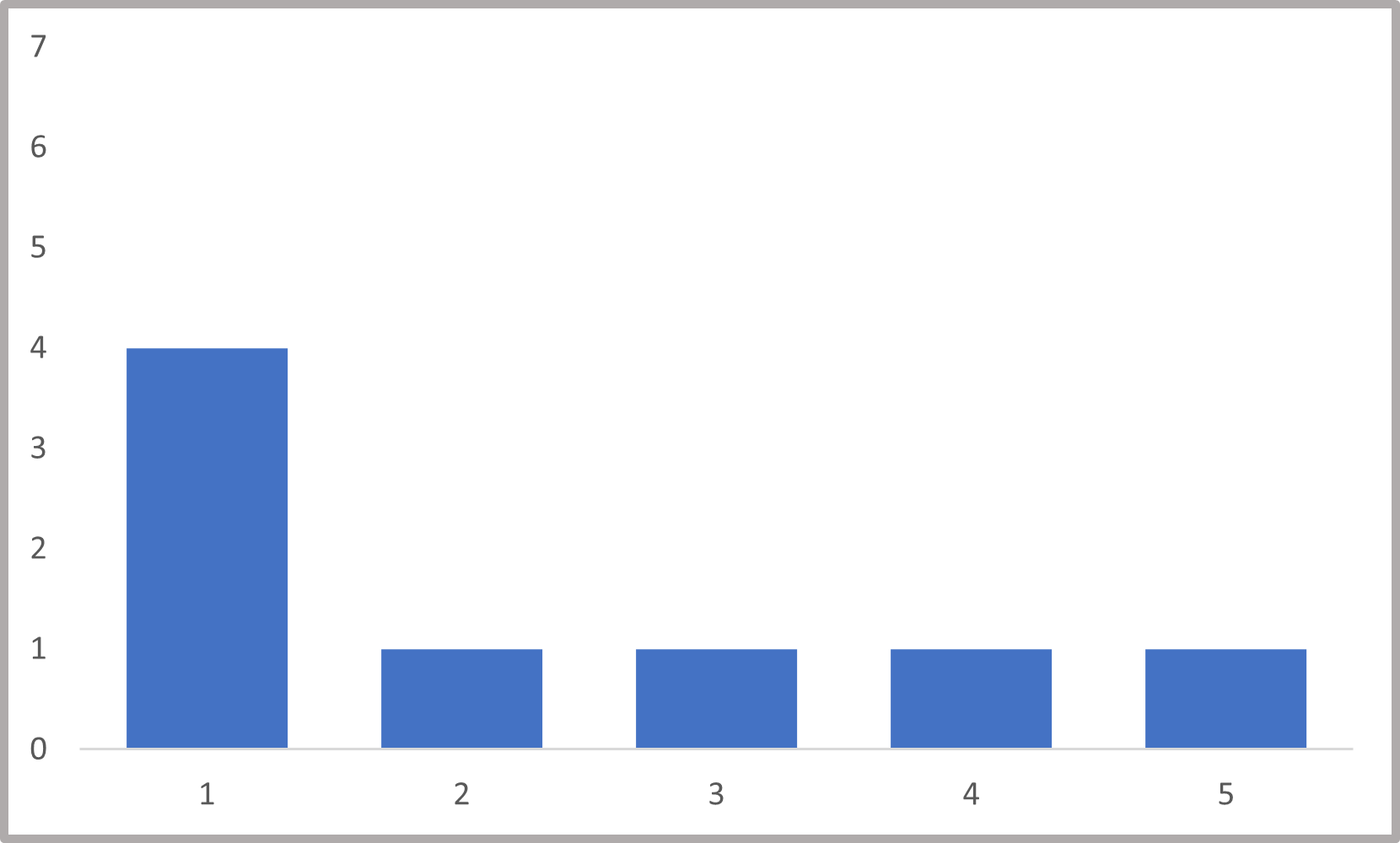}\\[-1pt]{2.3\tiny±1.6}}
 & \makecell{\includegraphics[width=0.8cm]{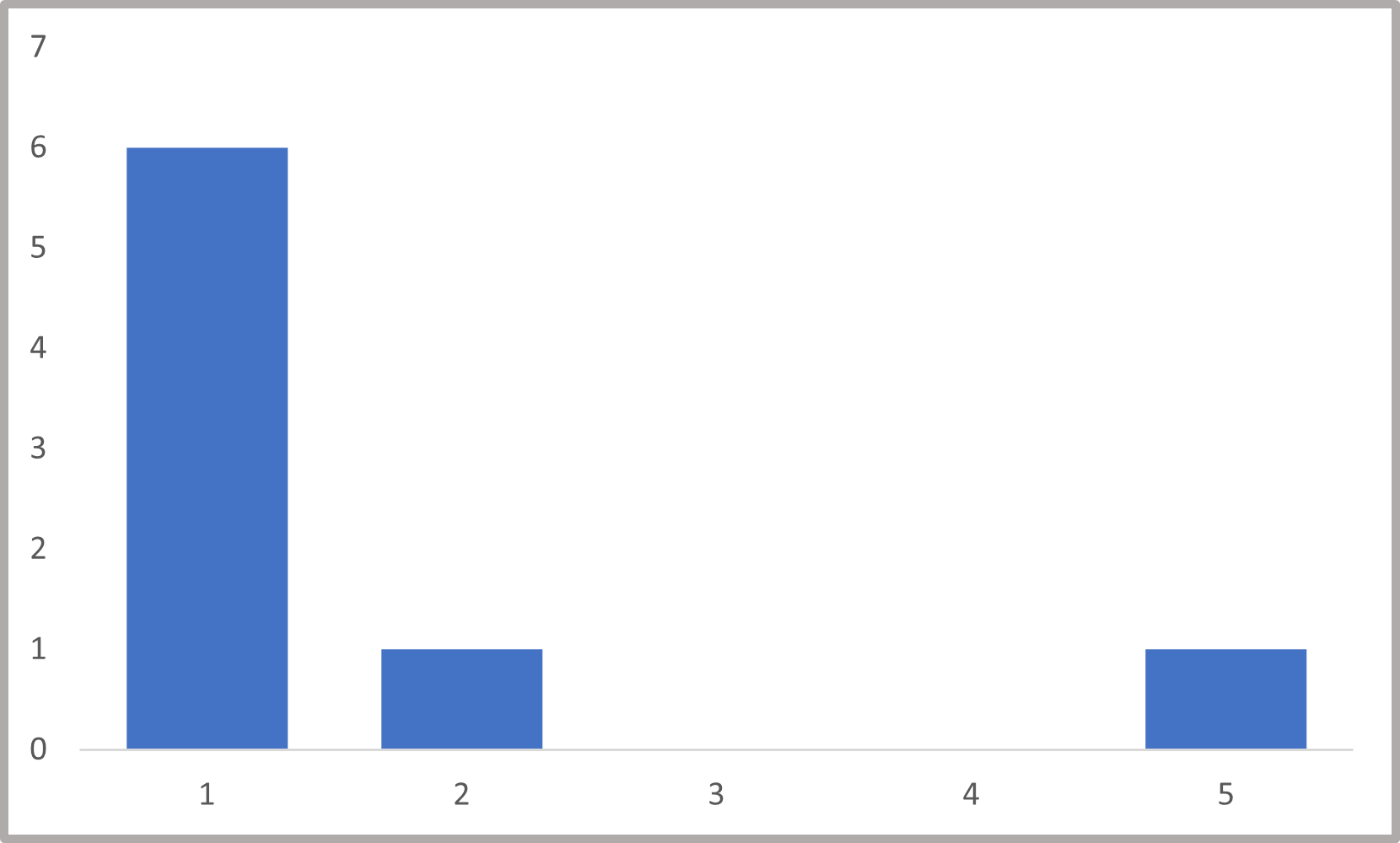}\\[-1pt]{\textcolor{orange}{1.6}\tiny±1.4}} \\
\cline{2-5}
 & \textbf{Automatic Failure Termination}: Automatically terminate execution and notify the user to take over when faults or unexpected states occur. 
 & \makecell{\includegraphics[width=0.8cm]{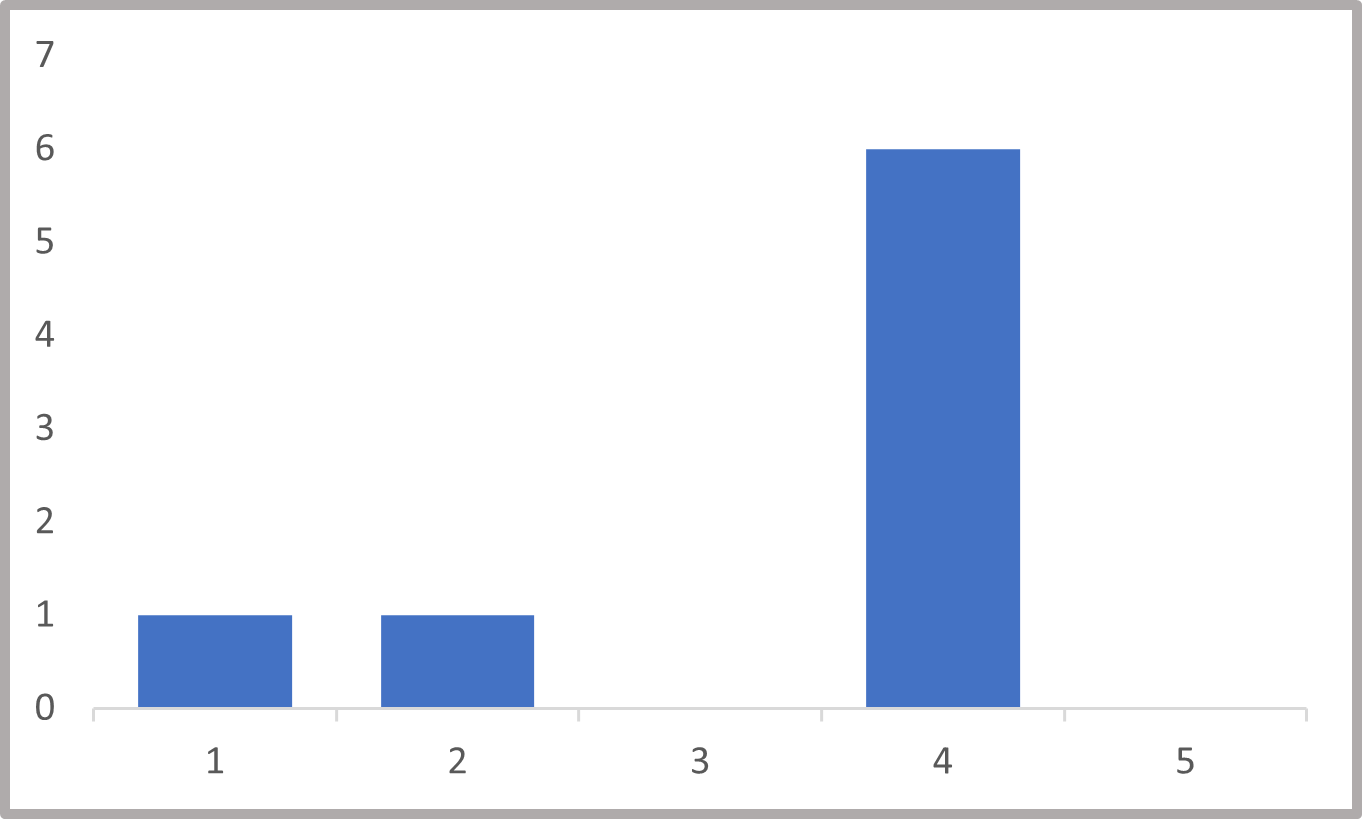}\\[-1pt]{3.4\tiny±1.2}}
 & \makecell{\includegraphics[width=0.8cm]{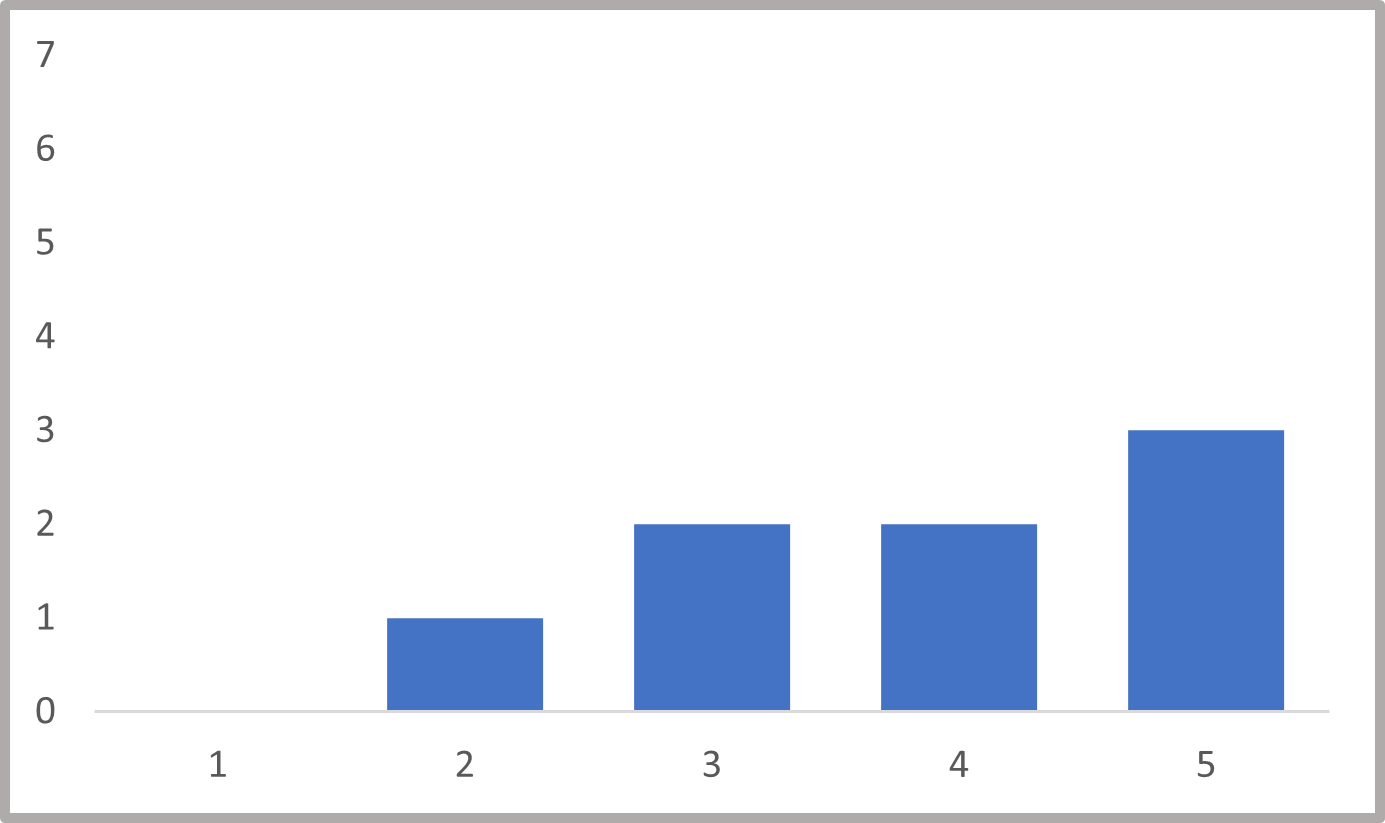}\\[-1pt]{3.9\tiny±1.1}}
 & \makecell{\includegraphics[width=0.8cm]{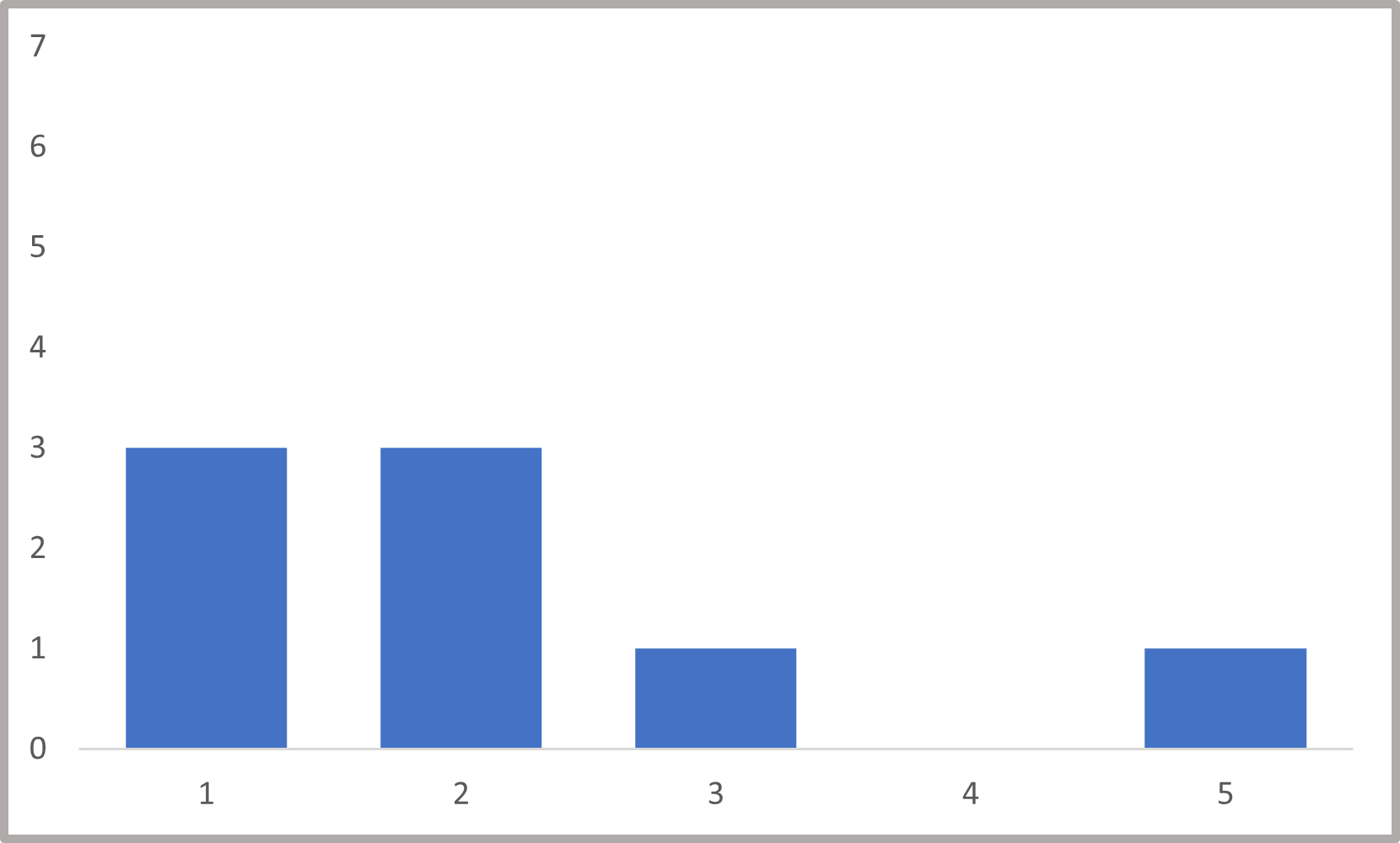}\\[-1pt]{2.1\tiny±1.4}} \\
\cline{2-5}
 & \textbf{High-Risk Permission Restrictions}: Restrict access to high-risk operations or permissions for vulnerable user groups. 
 & \makecell{\includegraphics[width=0.8cm]{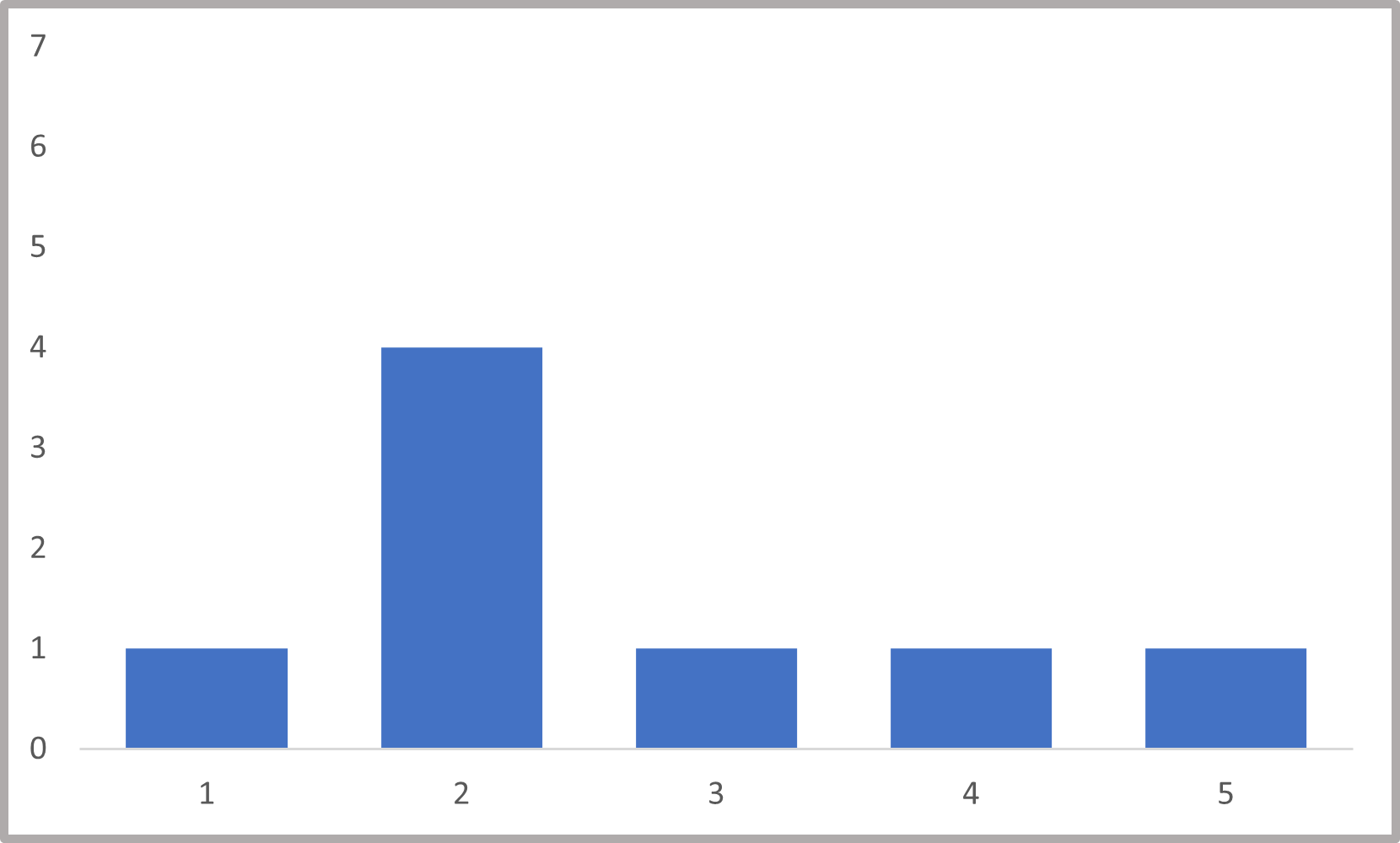}\\[-1pt]{2.6\tiny±1.3}}
 & \makecell{\includegraphics[width=0.8cm]{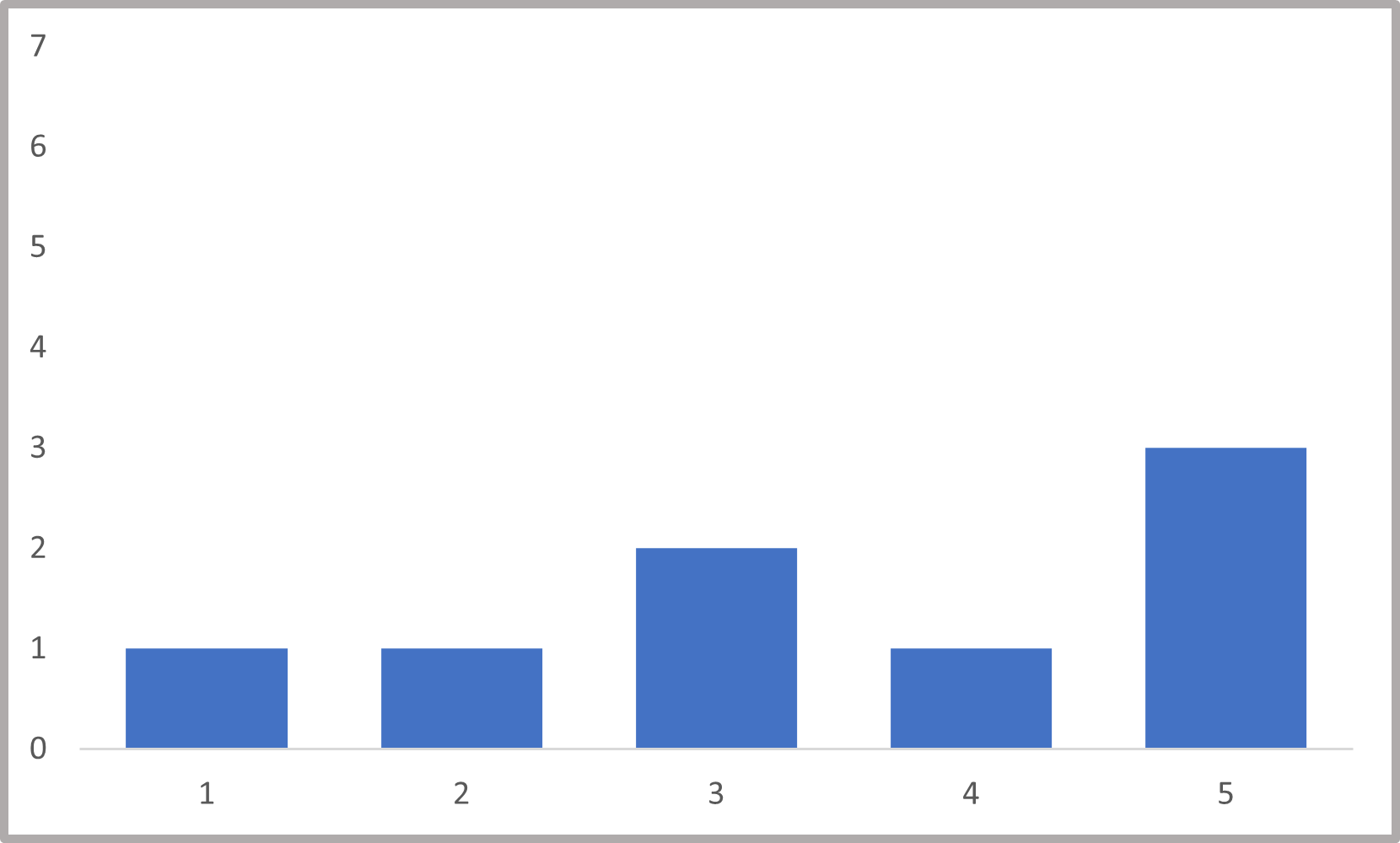}\\[-1pt]{3.5\tiny±1.5}}
 & \makecell{\includegraphics[width=0.8cm]{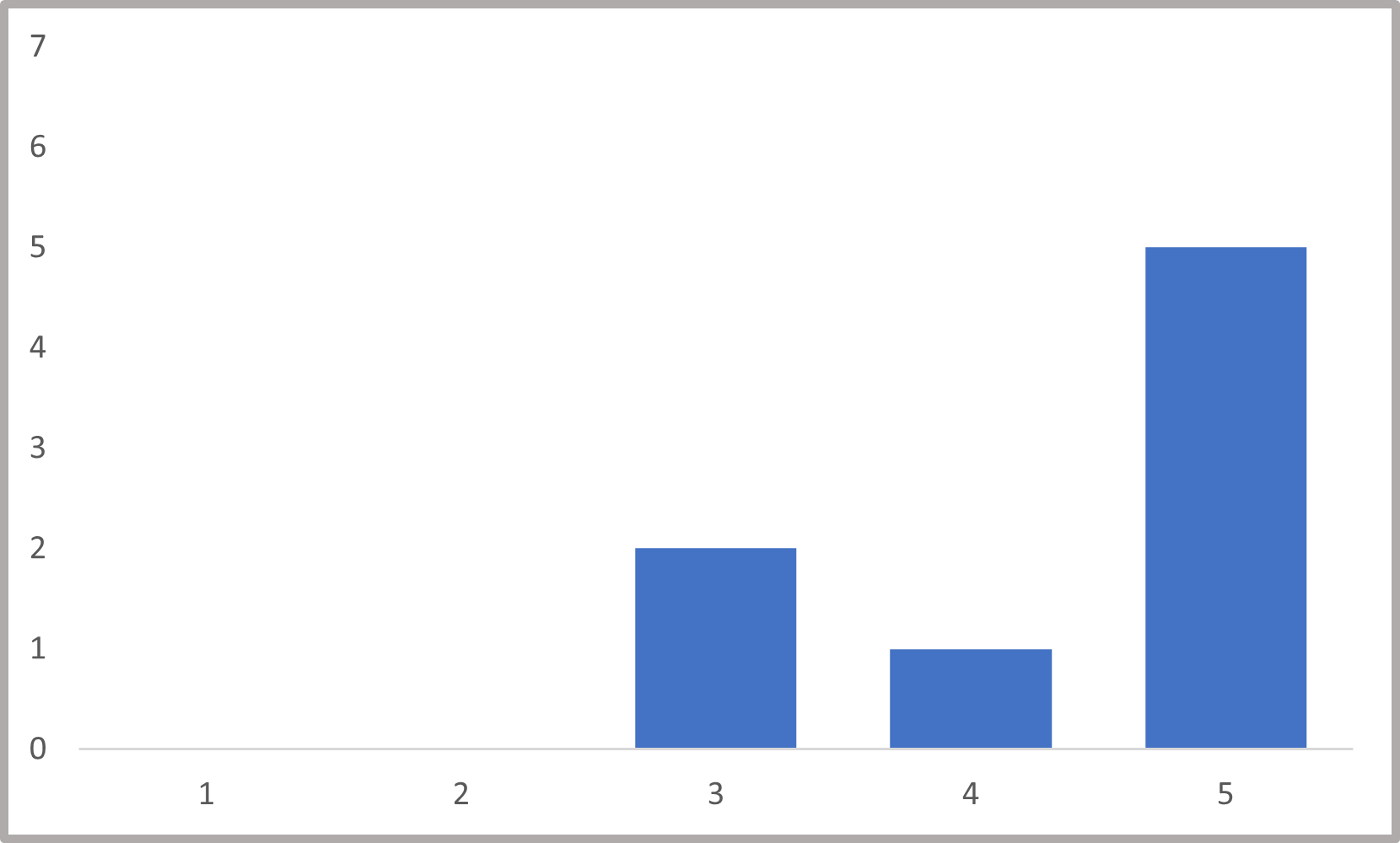}\\[-1pt]{\textcolor{green2}{4.4}\tiny±0.9}} \\
\hline
\multirow{6}{*}{Data Layer}
 & \textbf{Dedicated Data Management Space}: Provide a dedicated space for managing data uploaded to GenAI systems, allowing users to delete, modify, or encrypt their data. 
 & \makecell{\includegraphics[width=0.8cm]{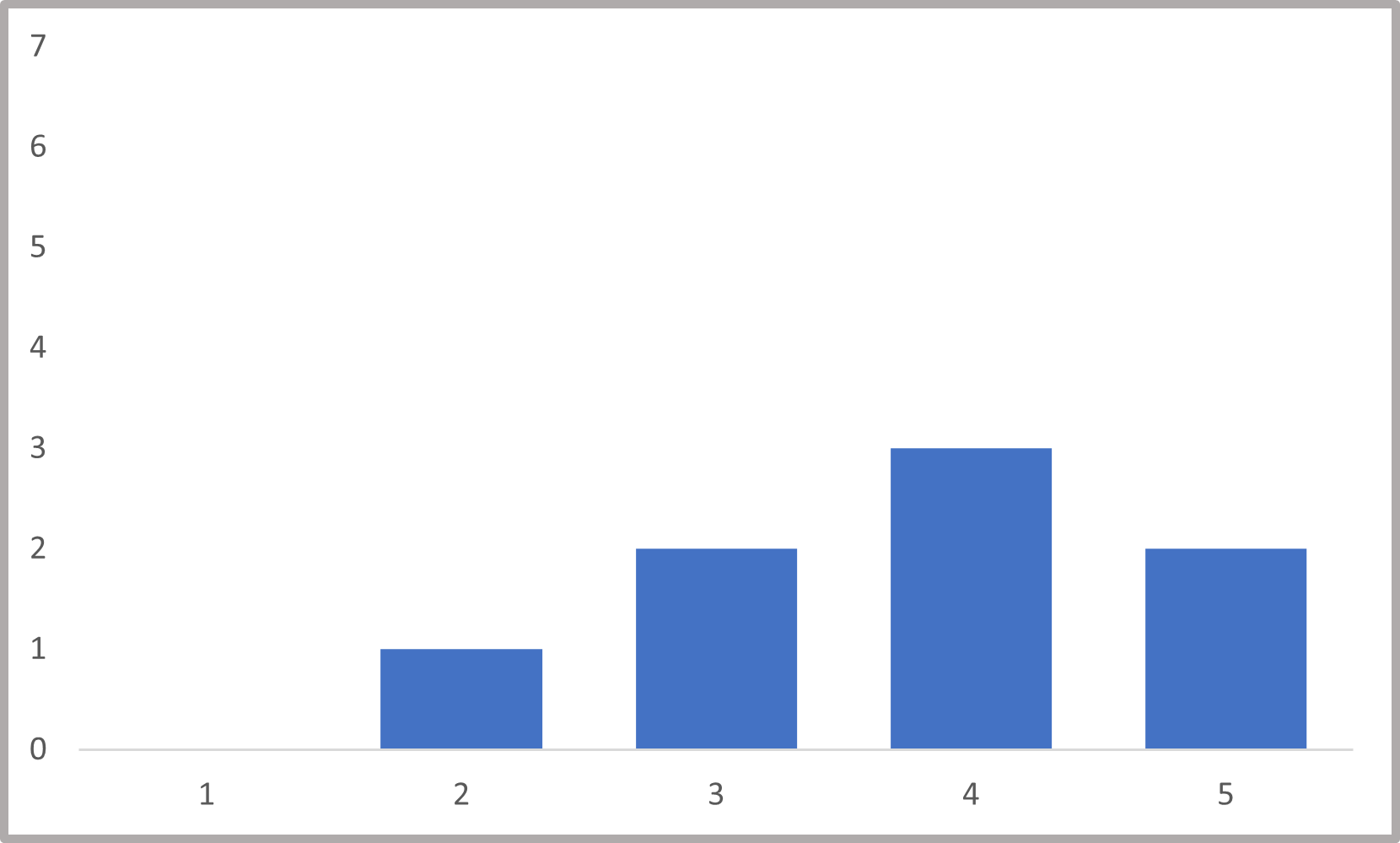}\\[-1pt]{3.8\tiny±1.0}}
 & \makecell{\includegraphics[width=0.8cm]{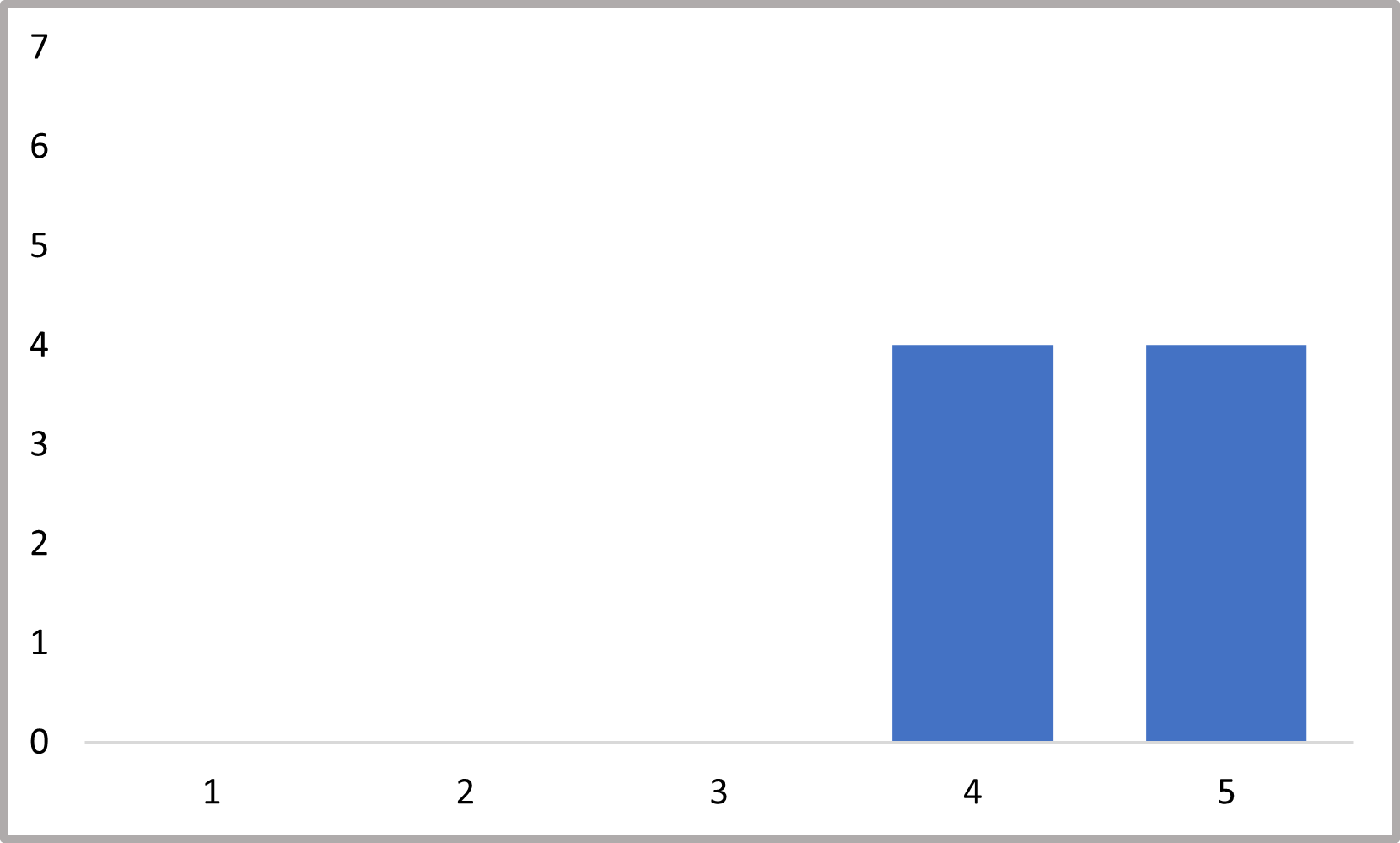}\\[-1pt]{\textcolor{green2}{4.5}\tiny±0.5}}
 & \makecell{\includegraphics[width=0.8cm]{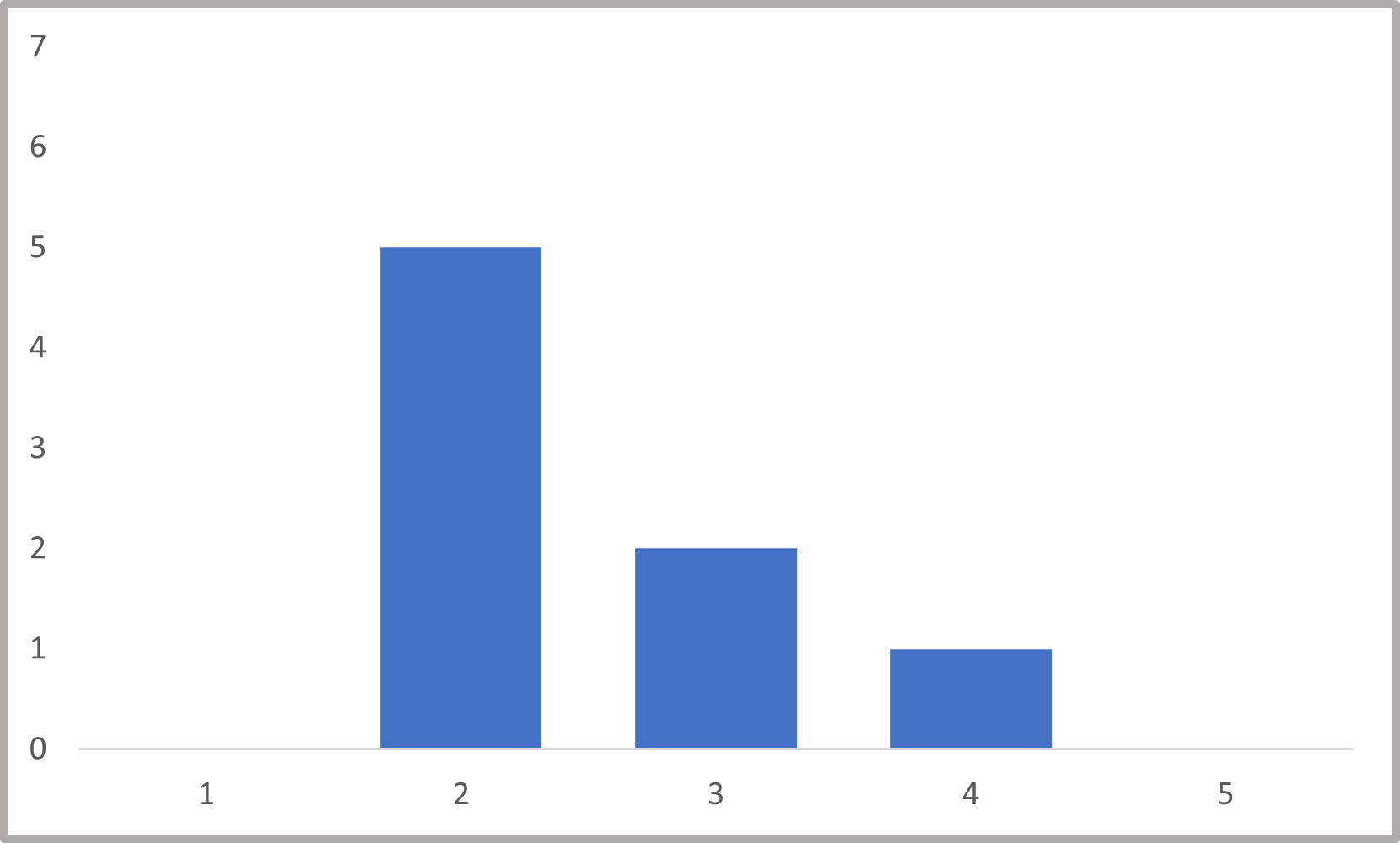}\\[-1pt]{3.5\tiny±0.8}} \\
\cline{2-5}
 & \textbf{Fine-Grained Data Control}: Enable fine-grained control over data access, such as allowing the assistant to access only selected photos. 
 & \makecell{\includegraphics[width=0.8cm]{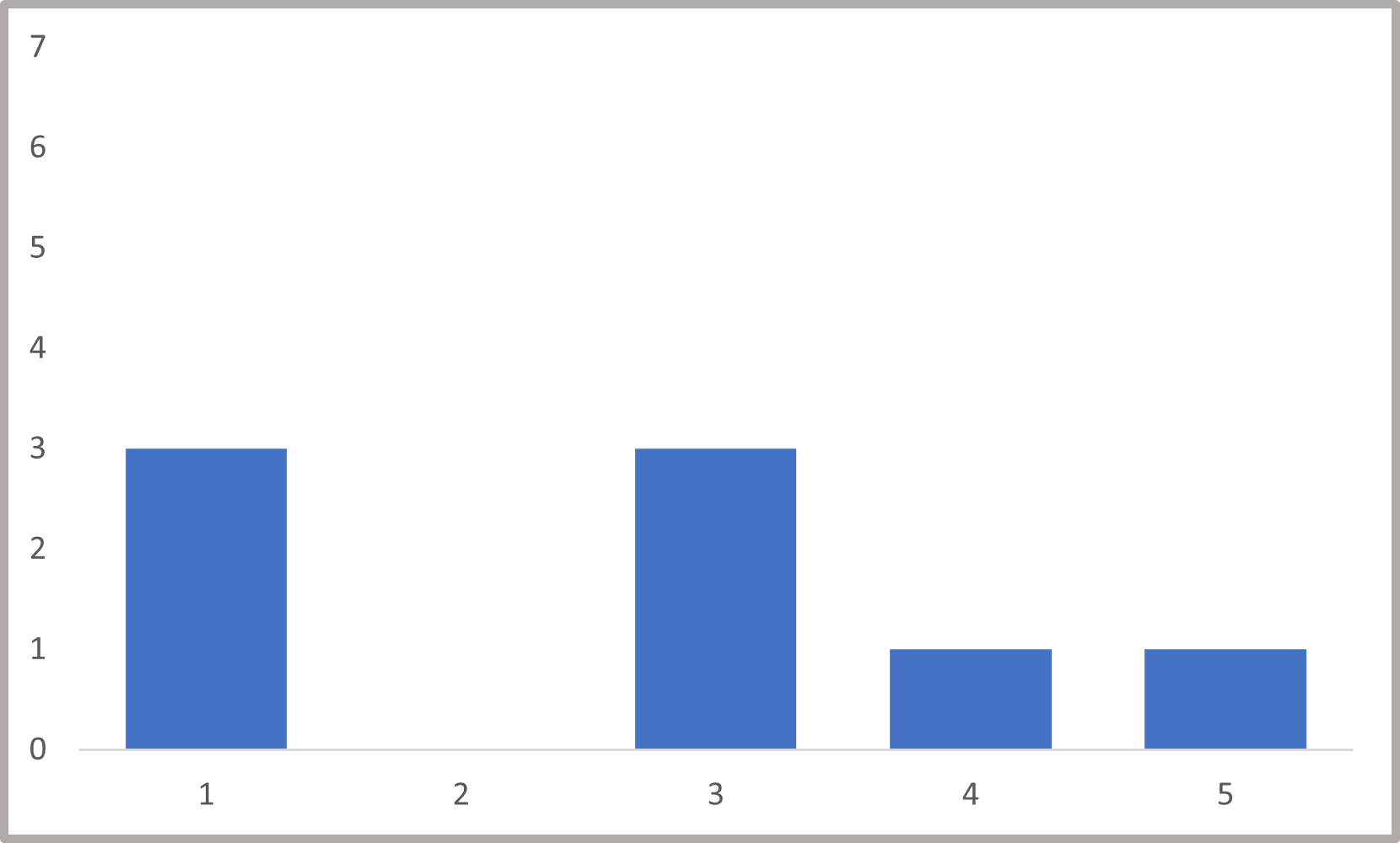}\\[-1pt]{2.6\tiny±1.5}}
 & \makecell{\includegraphics[width=0.8cm]{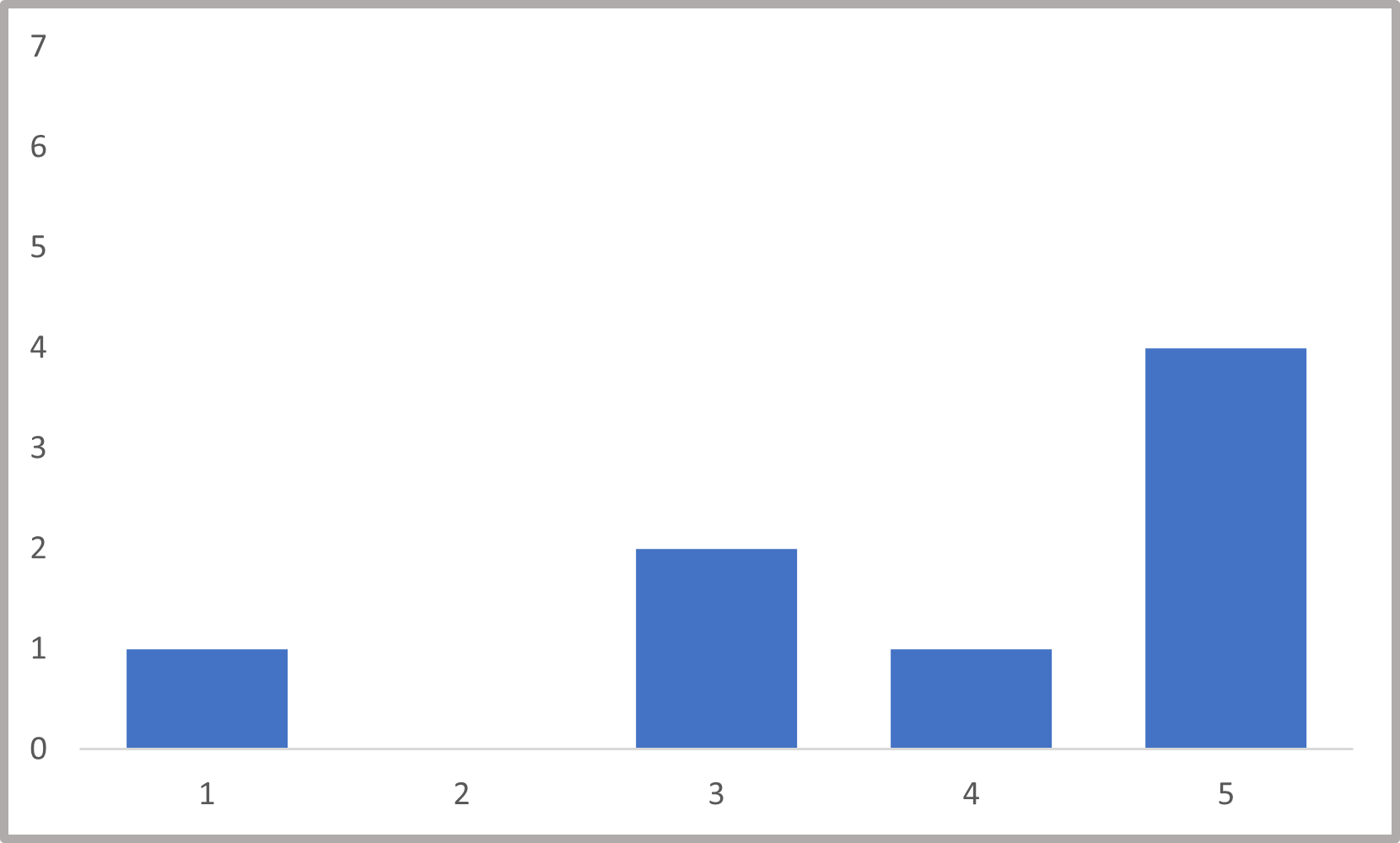}\\[-1pt]{3.9\tiny±1.5}}
 & \makecell{\includegraphics[width=0.8cm]{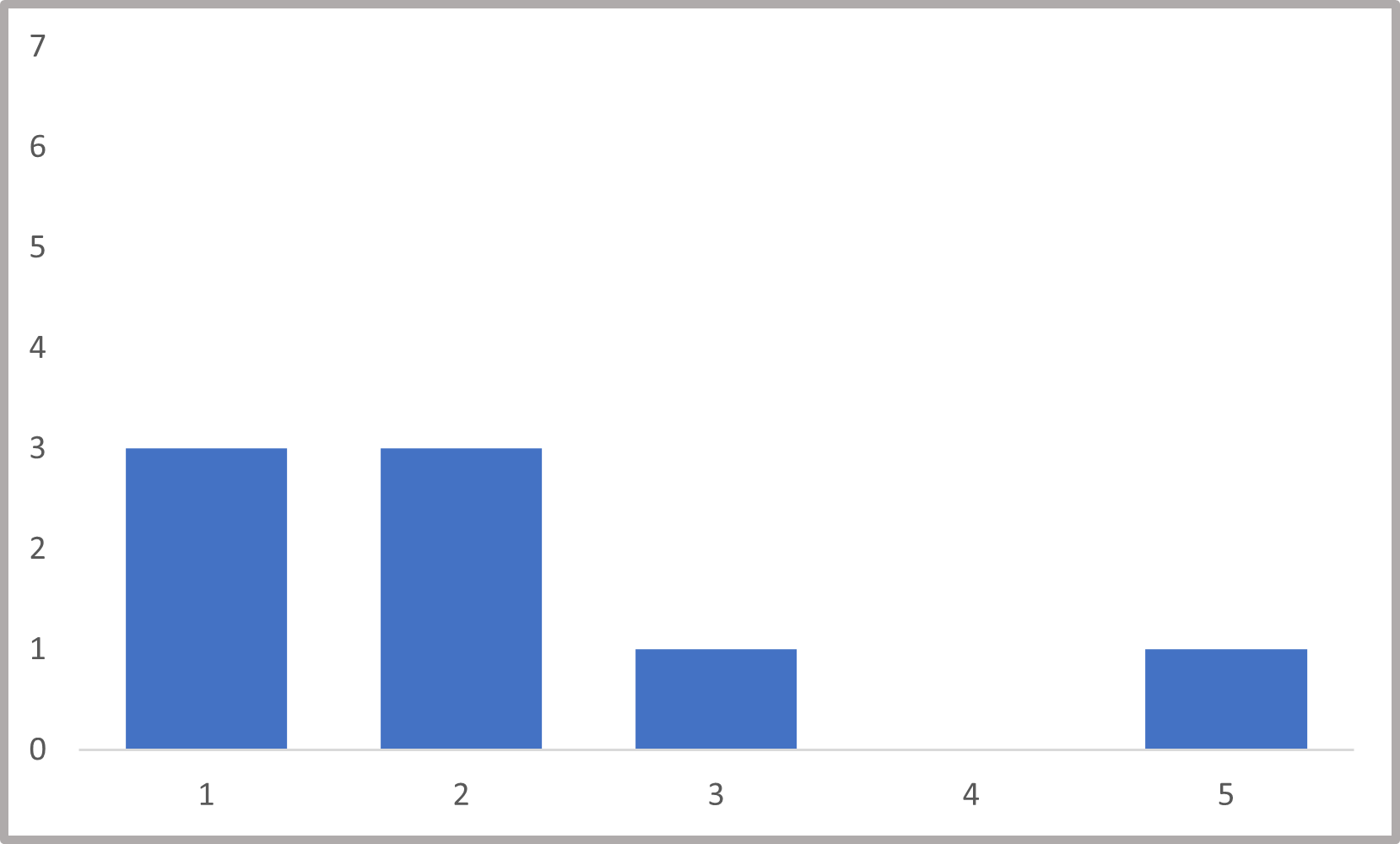}\\[-1pt]{3.9\tiny±1.4}} \\
\cline{2-5}
 & \textbf{Data Desensitization}: Identify sensitive information in collected data before uploading or storage, allowing users to choose which data should be anonymized or encrypted. 
 & \makecell{\includegraphics[width=0.8cm]{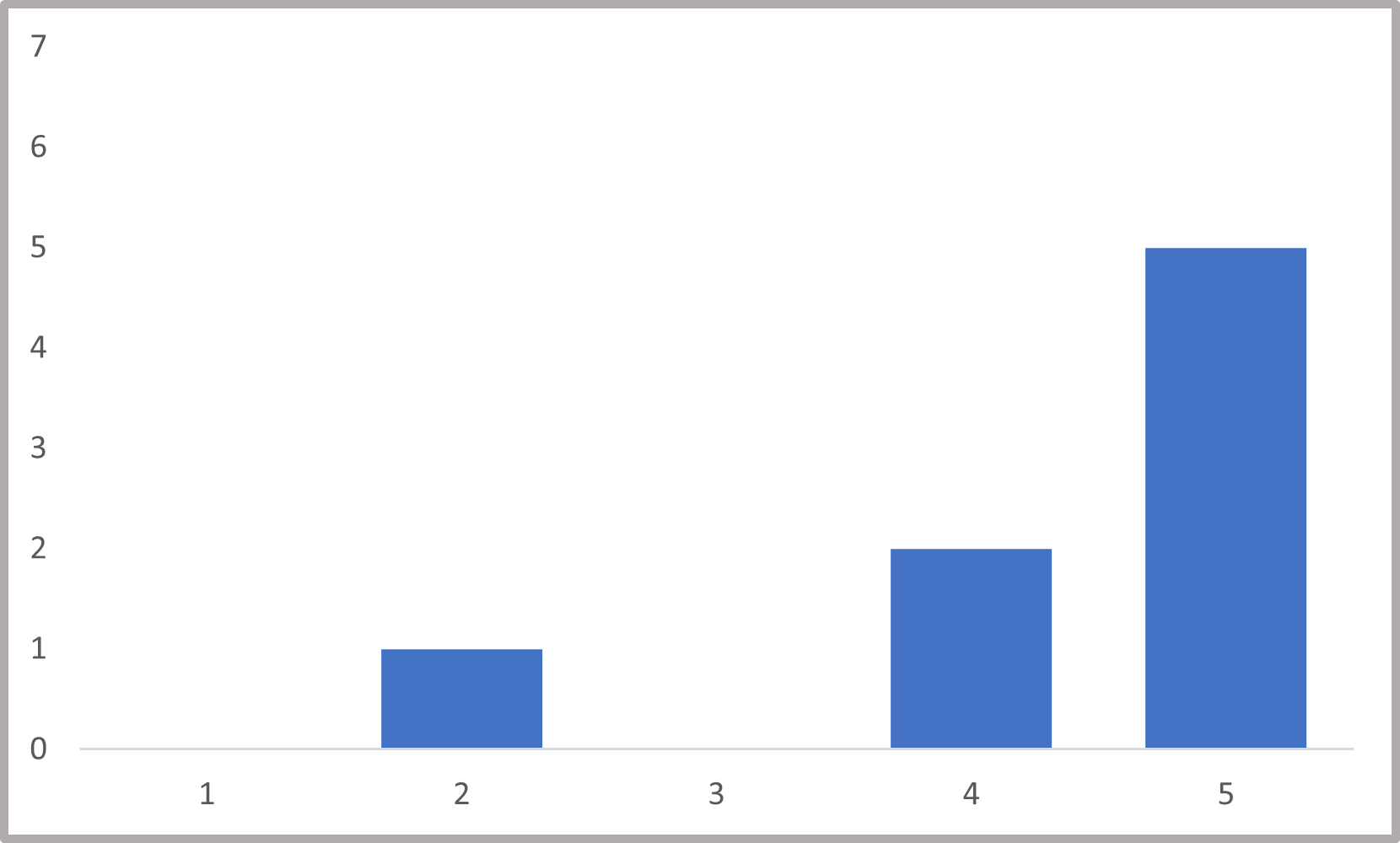}\\[-1pt]{\textcolor{green2}{4.4}\tiny±1.1}}
 & \makecell{\includegraphics[width=0.8cm]{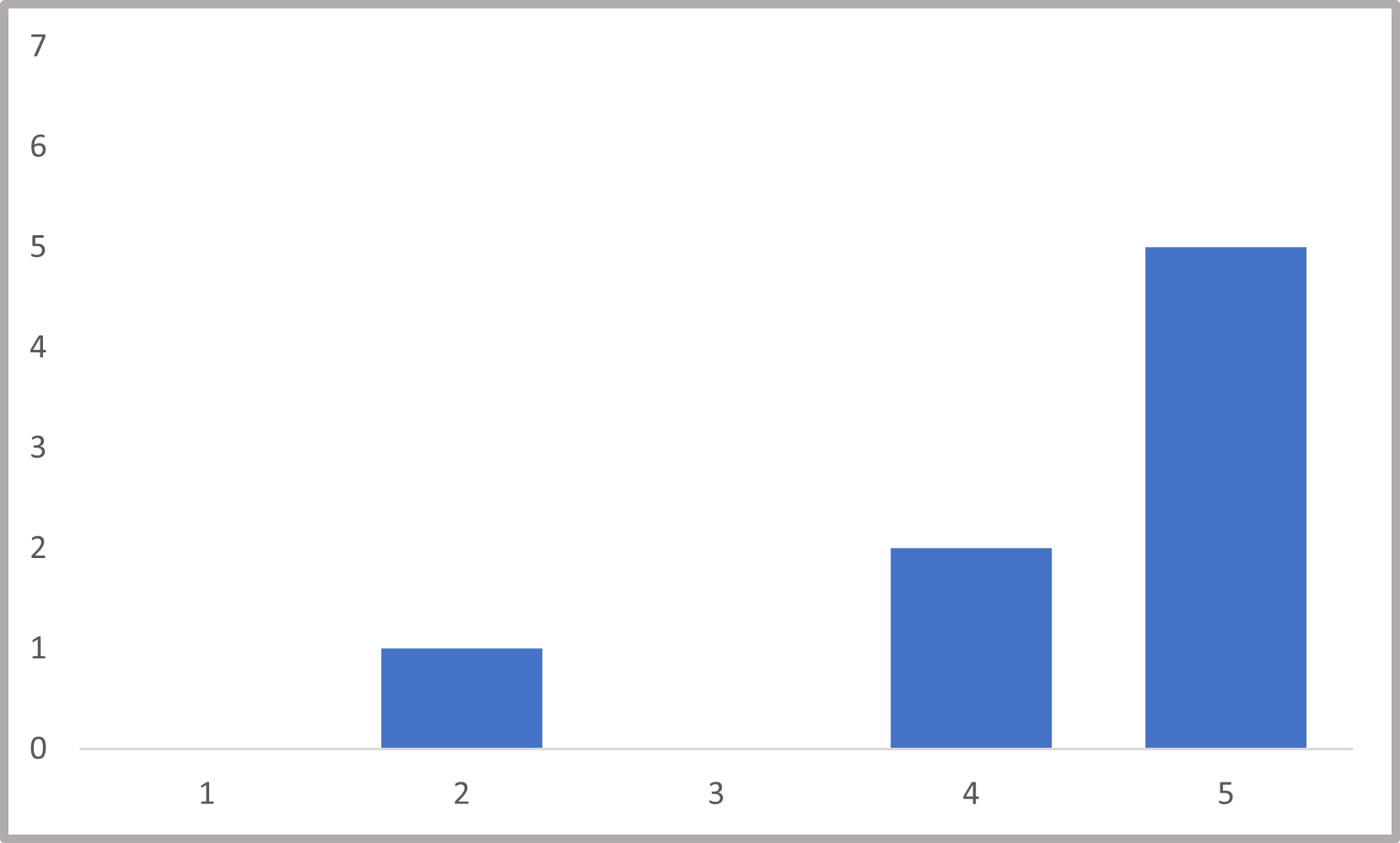}\\[-1pt]{\textcolor{green2}{4.4}\tiny ±1.1}}
 & \makecell{\includegraphics[width=0.8cm]{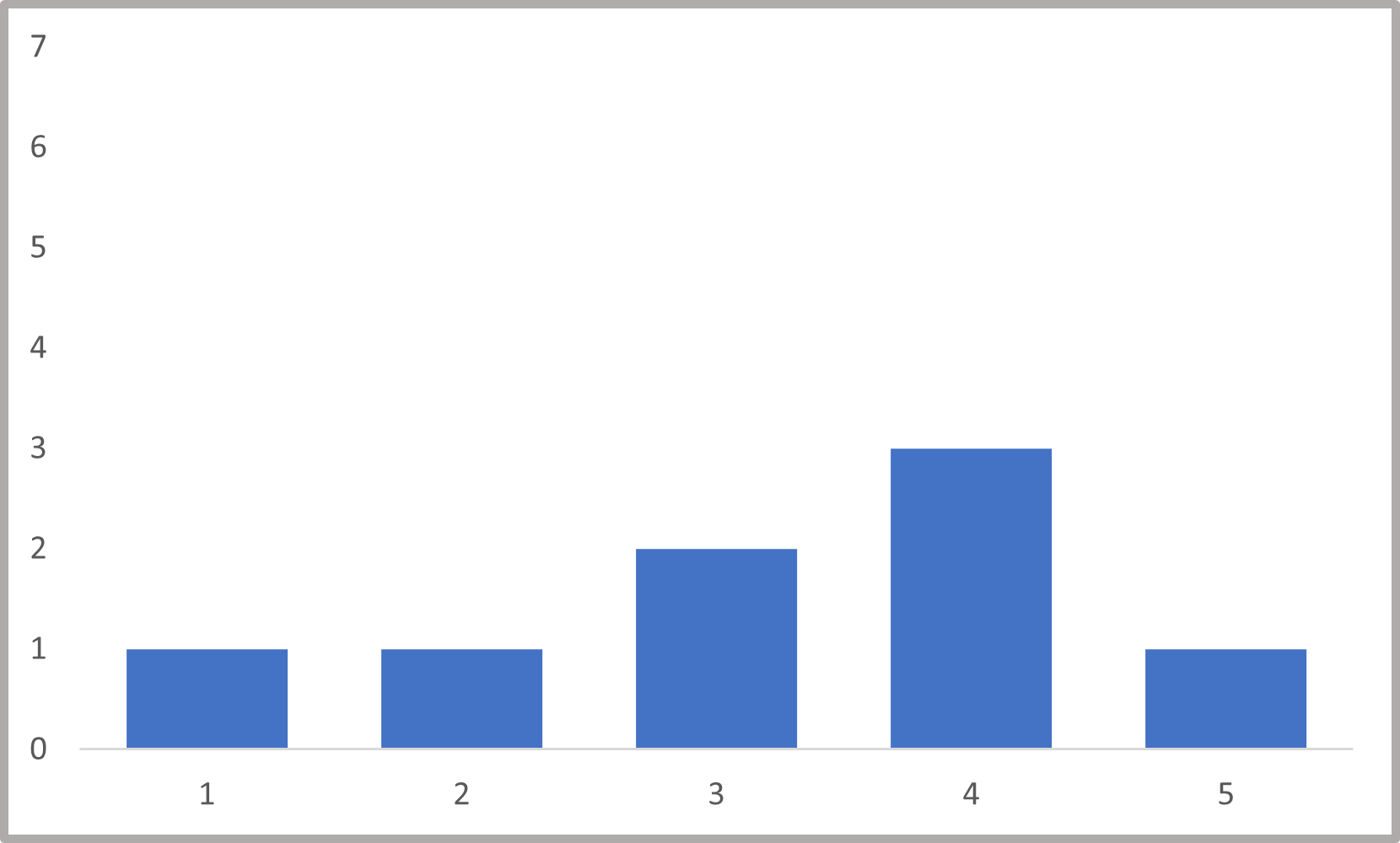}\\[-1pt]{2.8\tiny±1.3}} \\
\bottomrule
\end{tabularx}
\end{table*}
%
%

Based on the opinions we collected from interviews, we zoomed in on the system and data layer, and carefully summarized seven specific suggestions that practitioners could implement to enhance privacy in GenAI smartphones.
To understand the values of those suggestions, we further validated them through a focus group examining their effectiveness, feasibility, and implementation difficulty.
\autoref{tab:privacy_recommendations} summarizes suggestions and displays the score distribution of each dimension.
The mean scores of [\textbf{effectiveness}/ \textbf{feasibility}/ \textbf{difficulty}] are displayed in square brackets (e.g., \textbf{[\textcolor{green2}{5.0}/3.0/\textcolor{orange}{1.0}]}) after each item. We adjust the ``difficulty'' scale, making higher values indicate that the suggestion is technically easier to implement.

\subsubsection{System Layer}
We distilled four suggestions related to the design of permission framework and GenAI smarphone system.

\textbf{\textit{Permission Isolation.} [3.0/2.9/2.5] }
This suggestion proposes separating the permission management of native GenAI systems from that of all other post-installed apps (e.g., Instagram, Google Map), to effectively regulate GenAI-related operations.
Participants rated this suggestion as having relatively low effectiveness and feasibility. 
The low feasibility is a GenAI agents' core capabilities (e.g., reading on-screen content and simulating user interactions) are tightly coupled with existing accessibility permission in general app permission framework, making it difficult to achieve meaningful severance. 
Similarly, for effectiveness, many GenAI-related privacy risks arise from dynamic, context-driven behaviors during task execution, which cannot be fully constrained by static permission boundaries alone.
Implementation difficulty received polarized ratings, with three ratings at 2 and four ratings at 4–5. 
Participants who perceived greater difficulty emphasized that GenAI features operate through dynamic, context-dependent interactions, in which required permissions emerge at runtime rather than being statically defined, making isolation hard to pre-configure.

\textbf{\textit{Intent-Based Permission Framework.} [\textcolor{green2}{4.3}/2.3/\textcolor{orange}{1.6}]} This proposal advocates redesigning the permissions framework around user intents, rather than the existing data-centric permission framework. 
This intent-based permission framework aims to provide semantically meaningful access controls that align with how GenAI agents execute tasks and users to interpret. 
Scores indicate that the proposal is perceived as effective in mitigating privacy issues. 
Participants further envisioned a modular, script-based execution model, where complex tasks are decomposed into semantically clear and verifiable action units, similar to the \textit{skills} in modern AI agent scenario.
However, despite its conceptual appeal, participants highlighted substantial obstacles. 
Accurately decomposing ambiguous natural-language requests into stable, enforceable task steps was considered highly error-prone. Runtime enforcement would also require continuous monitoring to ensure execution remains aligned with declared intent.
Beyond technical challenges, ecosystem constraints further limit feasibility: requiring third-party apps to expose modular, intent-aware interfaces was viewed as unrealistic given existing business incentives. Overall, participants regarded intent-based permissions as a promising long-term direction rather than a near-term deployable solution.

\textbf{\textit{Automatic Failure Termination.} [3.4/3.9/2.1]} 
When the GenAI agents work under erroneous or unexpected conditions and anomalies are detected, the system would automatically stop execution and return control to users.  
Most participants considered it effective in mitigating risks arising from agent misbehavior (five ratings at 4), and feasibility was rated relatively high, with five participants assigning scores above 3. 
However, implementation was perceived as moderately to highly complex (six ratings at 4–5), reflecting concerns about reliably detecting fault conditions during real-time execution.
Although the feasibility is acceptable, participants noted that the core difficulty lies in determining abnormal execution. 
P1 noted: \textit{``A task may have multiple completion paths. How do we decide what constitutes an abnormal state? This seems quite challenging.''} 
GenAI agents operate across dynamic UI states and continuously changing contexts, making static definitions of failure impractical. Furthermore, participants warned that overly sensitive detection could cause frequent and disruptive interruptions, while conservative thresholds may fail to provide meaningful protection. Striking an effective balance between safety and usability therefore remains a key technical challenge.

\textbf{\textit{High-Risk Permission Restrictions.} [2.6/3.5/\textcolor{green2}{4.4}]} 
Drawing on \textit{parental control} models, this proposal suggests restricting GenAI systems' access to high-risk operations or sensitive permissions for vulnerable user groups such as children, older adults, or individuals with limited technical literacy. The aim is to introduce protective guardrails that prevent potentially harmful agent actions for users who may not fully understand the implications of GenAI operations. Participants expressed skepticism about the effectiveness of this approach in addressing privacy risks within GenAI smartphone contexts. Five participants arated its effectiveness as low (1–2), arguing that the core privacy challenges stem from the limited privacy awareness among certain user groups rather than the agent’s capabilities. They further noted that the convenience of GenAI smartphones primarily benefits users who are proficient with its technologies, and that directly restricting high-risk operations would significantly degrade user experience. Nevertheless, participants generally viewed the proposal as technically feasible and not presenting substantial implementation challenges.

\subsubsection{Data Layer}
We distilled three suggestions relatd to data management in GenAI smartphones. 

\textbf{\textit{Dedicated Data Management Space.} [3.8/\textcolor{green2}{4.5}/3.5]} 
To strengthen user control over personal data in GenAI systems, participants suggested providing a dedicated data management space where users can review, delete, modify, or encrypt data. 
Participants generally regarded this measure as moderately effective in addressing privacy concerns, with seven participants assigning scores of 3 or higher. 
They noted that the effectiveness of such a feature largely depends on users' trust in the system provider, particularly regarding whether data deletion and protection are actually enforced. Still, P6 stated, \textit{``Compared to third-party apps, first-party implementations are considered more reliable. Therefore, this design feels like a relatively effective and positive feature.''}
In terms of feasibility, the proposal was rated highly, with all participants assigning scores of 4-5, indicating strong technical and operational viability. Implementation difficulty was perceived as relatively low, with seven ratings of 3 or below. Overall, participants widely recognized the value of this suggestion and regarded its implementation as feasible. 

\textbf{\textit{Fine-Grained Data Control.} [2.6/3.9/3.9]} 
Participants suggested granular controls over data access, such as restricting assistants to specific photos, to support more precise personal data management. However, this approach was generally rated as having limited effectiveness, with six ratings of 3 or below. Several participants noted that item-by-item authorization would impose substantial interaction overhead. As P1 described it, the process would be ``excessively cumbersome,'' and P2 claimed that it would ``severely impact user experience,'' adding that existing photo picker mechanisms offer limited practical value~\cite{apple2025limited-photo-library, android2025photo-picker}. P4 further noted that granular control is not suitable for all data types: while structured data like photos or contacts can be segmented, unstructured inputs generated during execution (e.g., addresses or ID numbers) are difficult to meaningfully isolate. 
Despite these concerns, participants viewed the proposal as technically feasible. Five participants rated feasibility as high (4–5), and six rated implementation difficulty as low (1–2), indicating confidence in its practicality. 
To mitigate usability issues, participants further suggested complementing granular controls with \textit{data classification schemes} that group user data into multiple sensitivity tiers. Under this scheme, low-sensitivity data (e.g., usernames or public profile) could be accessed with minimal friction after authorization, whereas high-sensitivity data (e.g., ID numbers, bank accounts, biometrics) would be subject to stricter controls.

\textbf{\textit{Data Desensitization.} [\textcolor{green2}{4.4}/\textcolor{green2}{4.4}/2.8]} 
Participants proposed applying data desensitization mechanisms to identify sensitive information before data is uploaded or stored, enabling selective anonymization or encryption. This approach was viewed as highly effective in addressing privacy concerns, with seven participants assigning scores of 4–5. 
Feasibility was also rated positively, as most participants believed such mechanisms align well with existing technical capabilities. 
Whereas assessments of implementation difficulty were more mixed. 
Some noted that identifying sensitive information is challenging due to the diversity of tasks and data types. As P2 stated, \textit{``I think the hardest part is actually determining whether information is sensitive and how to cover all tasks.''} 
In contrast, P4 argued that identification could be manageable, pointing out that regulations already define common categories of sensitive data such as addresses, phone numbers, and ID numbers. P8 further observed that existing detection systems, while imperfect, can achieve reasonably high accuracy in practice. 

\section{Related Work}
\label{related work}
The widespread adoption of GenAI has triggered growing privacy concerns. 
A growing body of research has investigated users’ perceptions of GenAI-related privacy risks across 
domains such as AR/VR/XR~\cite{alkaeed2024privacy, abraham2022implications, o2023privacy} and smart home systems~\cite{zeng2017end, marky2021roles}, as well as in general AI apps~\cite{zufferey2025ai, zhang2024s,Acquisti2020SecretsAL,jung2024we}.
These studies consistently highlight user concerns about the transparency of data collection, storage, and usage in AI systems~\cite{Shklovski2014Leakiness, Binns2018ItsRA, leschanowsky2024evaluating}. 
While users often appreciate the convenience afforded by GenAI, they remain apprehensive about unauthorized access, continuous background data collection, and potential vulnerabilities and attacks targeting GenAI models~\cite{Lau2018Alexa,Swetha2025PrivacyPI,Li2023Privacy}. 
For example, Li et al.~\cite{Li2023Privacy} provided a comprehensive survey of privacy attacks and defenses in GenAI models, categorizing threats such as extraction, inversion, and membership inference attacks.
As GenAI capabilities increasingly migrate to mobile platforms~\cite{Xue2024PowerInfer2FL, Lin2023AWQAW, Semerikov2025LLMOT}, Wu et al.~\cite{wu2024first} were among the first to systematically investigate the current state and challenges of integrating LLMs into smartphones, highlighting the risks of privacy leakage and data security threats in mobile GUI agents.
Our work extends this line of privacy research by focusing on GenAI smartphones, where GenAI is integrated at the system level rather than confined to individual apps. To our knowledge, this study presents the first in-depth, user-centered investigation of privacy perceptions and concerns in this emerging context,
providing timely insights to inform privacy-aware design at this early adoption stage.

\section{Conclusion}
\label{conclusion}
This study provides an early empirical exploration of user privacy perspectives on GenAI smartphones. Through interviews with 22 participants, we find that users' privacy concerns often intensify once they gain awareness of the underlying system mechanisms. These concerns span opaque data collection, unpredictable AI behavior, insufficient regulatory safeguards, and security vulnerabilities.
Participants proposed privacy-enhancing suggestions across three layers: system-level control mechanisms, data lifecycle management, and user-facing transparency.
Our findings provide timely user-centered evidence to inform the design of more privacy-aware and user-friendly GenAI smartphones.

\appendix
\section*{Ethical Considerations}
This study adhered to established ethical standards for research involving human participants. All participants were adults who voluntarily agreed to participate in the study. Before each interview or survey, participants were clearly informed about the study’s purpose, procedures, and potential risks. Written or electronic informed consent was obtained, and participants were explicitly informed that their participation was entirely voluntary and that they could withdraw at any time without penalty.
To protect participants’ privacy, the interview content was kept strictly confidential and accessible only to the research team members involved in transcription and follow-up communication. 
In all transcripts and analyzes, identifying information was removed or anonymized. 
Data were securely stored on password-protected servers, and no personal information will be disclosed to anyone outside the research team. 
The study received approval from the Institutional Review Board (IRB). All data will be used strictly for academic purposes and will not be shared with third parties in an identifiable form.
\section*{Open Science}
To enhance transparency, reproducibility, and meta-research, and to comply with open science policies, we provide the following research artifacts in the online supplementary materials~\cite{interviews-supplemental-materials}: (1) the preliminary learning document, (2) interview questions, (3) the final codebook, and (4) the focus group discussion protocol. In accordance with data protection requirements and ethical research practices, the original interview data, including audio recordings and transcripts, are not included in the replication package. This reflects our commitment to protecting participants’ privacy and safeguarding their data rights. By doing so, we mitigate the risk of disclosing any information that could potentially identify our participants or their projects, whether through contextual details or metadata. Instead, we share our research findings using thematic analysis and anonymized interview excerpts.

\bibliographystyle{ACM-Reference-Format}
\bibliography{base}
\appendix
\pagebreak
\section{Demographic information of participants}
\begin{table}[!h]
\centering
\caption{Demographic information of participants (n = 22).}
\label{tab:participants}
\resizebox{0.98\linewidth}{!}{%
\begin{tabular}{l l c c}
\toprule
\textbf{Category} & \textbf{Group} & \textbf{Count} & \textbf{Percentage} \\
\midrule
Gender     & Male        & 11 & 50\% \\
           & Female      & 11 & 50\% \\
\midrule

Education Degree  & Associate   & 1  & 5\%  \\
           & Bachelor    & 13 & 59\% \\
           & Master      & 6  & 27\% \\
           & Ph.D.       & 2  & 9\%  \\
           
\midrule
IT/CS Background      & Yes     & 13 & 59\% \\
                      & No   & 9 & 41\% \\
\midrule
GenAI Smartphone Use  & Current/Former users  & 15 & 68\% \\
           & Tried others’ devices  & 1  & 5\%  \\
           & Never used             & 6  & 27\% \\
\midrule
Smartphone OS         & Android     & 14 & 64\% \\
           & iOS         & 8  & 36\% \\
\midrule
Understanding Level of & None   & 1 & 5\%\\
              GenAI Smartphone & Basic & 10 & 45\% \\
              & Intermediate  & 9  & 41\% \\
              & Extensive    & 2  & 9\%  \\
\bottomrule
\end{tabular}
}%
\end{table}

\end{document}